\documentclass[aps,prl,reprint, superscriptaddress]{revtex4-2}
\usepackage{amsmath}
\usepackage{amsfonts}
\usepackage{amssymb}
\usepackage{graphicx}
\usepackage{dcolumn}
\usepackage{bm}
\usepackage{hyperref}
\usepackage{changes}%
\usepackage{placeins}
\usepackage{braket}
\newcommand{\angstrom}{\mbox{\normalfont\AA}}
\newcommand{\Rnl}{\ensuremath{R_\mathrm{nl}}}
\newcommand{\Rsq}{\ensuremath{R_\mathrm{sq}}}
\newcommand{\Idc}{\ensuremath{I_\mathrm{DC}}}
\newcommand{\Idelta}{\ensuremath{I_{\delta}}}
\newcommand{\DRnl}{\ensuremath{\Delta R_\mathrm{nl}}}

\newcommand{\Vbg}{\ensuremath{V_\mathrm{bg}}}
\newcommand{\tpar}{\ensuremath{\tau_s^{\parallel}}}
\newcommand{\Ds}{\ensuremath{D_s}}
\newcommand{\tp}{\ensuremath{\tau_p}}

\newcommand{\tso}{\ensuremath{\tau_\mathrm{so}}}
\newcommand{\tiv}{\ensuremath{\tau_\mathrm{iv}}}
\newcommand{\tvz}{\ensuremath{\tau_\mathrm{VZ}}}
\newcommand{\lvz}{\ensuremath{\lambda_\mathrm{VZ}}}
\newcommand{\tper}{\ensuremath{\tau_s^{\perp}}}

\newcommand{\Dc}{\ensuremath{D_c}}
\newcommand{\Vcnp}{\ensuremath{V_\mathrm{cnp}}}
\newcommand{\Rgr}{\ensuremath{R_\mathrm{sq}^\mathrm{gr}}}
\newcommand{\RgrTMD}{\ensuremath{R_\mathrm{sq}^\mathrm{gr/TMD}}}
\newcommand{\liA}[1]{\ensuremath{\lambda_\mathrm{I}^\mathrm{A#1}}}
\newcommand{\liB}[1]{\ensuremath{\lambda_\mathrm{I}^\mathrm{B#1}}}
\newcommand{\lo}{\ensuremath{\lambda_\mathrm{0}}}
\newcommand{\lr}{\ensuremath{\lambda_\mathrm{R}}}
\newcommand{\musy}{\ensuremath{\mu_{sy}}}
\newcommand{\g}[1]{\gamma_{#1}}

\begin{document}

\title{Electrical control of valley-Zeeman spin-orbit coupling-induced spin precession at room temperature}

\author{Josep Ingla-Ayn\'es}
\email{j.ingla@nanogune.eu}
\affiliation{CIC nanoGUNE BRTA, 20018 Donostia-San Sebastian, Basque Country, Spain}
\author{Franz Herling}
\affiliation{CIC nanoGUNE BRTA, 20018 Donostia-San Sebastian, Basque Country, Spain}
\author{Jaroslav Fabian}
\affiliation{Institute for Theoretical Physics, University of Regensburg, 93040 Regensburg, Germany}
\author{Luis E. Hueso}
\affiliation{CIC nanoGUNE BRTA, 20018 Donostia-San Sebastian, Basque Country, Spain}
\affiliation{IKERBASQUE, Basque Foundation for Science, 48013 Bilbao, Basque Country, Spain}
\author{F\`elix Casanova}
\email{f.casanova@nanogune.eu}
\affiliation{CIC nanoGUNE BRTA, 20018 Donostia-San Sebastian, Basque Country, Spain}
\affiliation{IKERBASQUE, Basque Foundation for Science, 48013 Bilbao, Basque Country, Spain}
\date{\today}
\begin{abstract}
The ultimate goal of spintronics is achieving electrically controlled coherent manipulation of the electron spin at room temperature to enable devices such as spin field-effect transistors. With conventional materials, coherent spin precession has been observed in the ballistic regime and at low temperatures only. However, the strong spin anisotropy and the valley character of the electronic states in 2D materials provide unique control knobs to manipulate spin precession. Here, by manipulating the anisotropic spin-orbit coupling in bilayer graphene by the proximity effect to WSe$_2$, we achieve coherent spin precession in the absence of an external magnetic field, even in the diffusive regime. Remarkably, the sign of the precessing spin polarization  can be tuned by a back gate voltage and by a drift current. Our realization of a spin field-effect transistor at room temperature is a cornerstone for the implementation of energy efficient spin-based logic.
\end{abstract}
\maketitle


The realization of logic operations using the spin degree of freedom is a crucial goal for spintronics \cite{dery2007,behin2010,manipatruni2019,pham2020,dieny2020}.
In this context, one of the most studied theoretical proposals is that of Datta and Das \cite{DattaDas}, which requires spin precession around the spin-orbit fields (SOFs) and has raised considerable interest \cite{dieny2020,DattaDas,zutic2004,schliemann2003,liu2011}. However, the experimental achievement of the required strong spin-orbit coupling (SOC) regime in conventional materials can only be realized in ballistic systems with long momentum scattering time (\tp{}) and very clean interfaces \cite{zutic2004,schliemann2003,liu2011}. Consequently, its implementation in all-electrical devices is currently limited to low temperatures \cite{JohnsonDattaDas,
wunderlich2010,olejnik2012,SpinTransistor2015,choi2015,choi2018}. 

\begin{figure}[tb]
	\centering
		\includegraphics[width=0.5\textwidth]{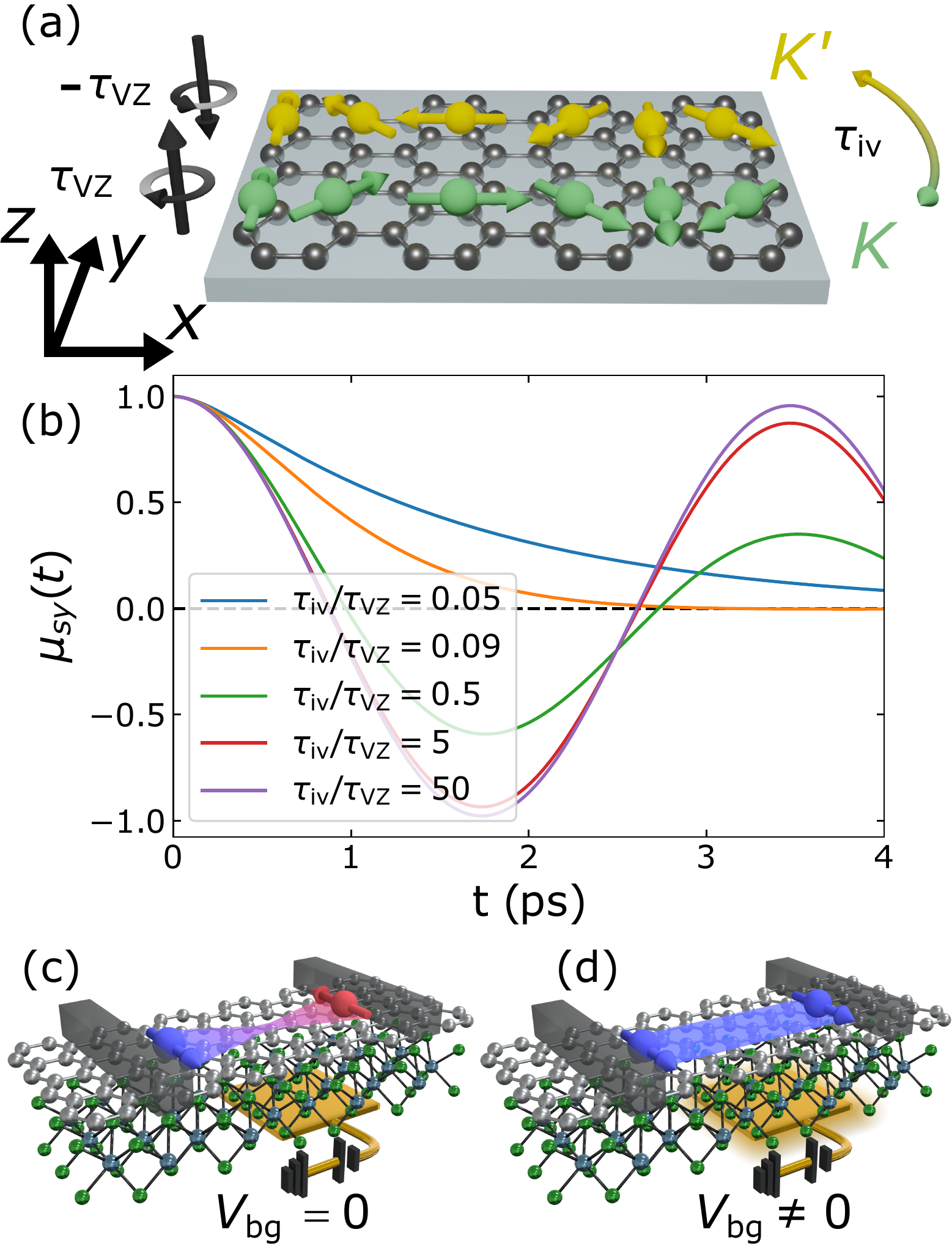}
	\caption{{Device working principle and BLG/WSe$_2$ spin transistor operation.} (a) Sketch of a
BLG/WSe$_2$ heterostructure. Out-of-plane valley-Zeeman SOF (black arrows) with opposite sign at the K
and K' valleys induce in-plane spin precession with a period \tvz{}. Spins can scatter between the
valleys via intervalley scattering (\tiv{}). (b) Time dependence of the spin accumulation $\mu_{sy}$ for different \tiv{} values (see Ref.~\cite{supinfo} for details). $\mu_{sy}$ undergoes net precession for \tiv{}$\geq$0.5\tvz{}. (c) and (d) Sketch of a spin field-effect transistor operating at the strong SOC regime where the valley-Zeeman induced spin precession is tuned by \Vbg{} to control the sign reversal.
	} 
	\label{Figure1}
\end{figure}

Alternatively, graphene-based van der Waals heterostructures are an ideal platform for spin manipulation \cite{reviewFabian, garcia2018} since, in these systems, graphene's low SOC can be enhanced by proximity with transition metal dichalcogenides (TMDs) \cite{gmitra2016,wang2016,yang2016,zihlmann2018,wakamura2018,cummings2017,ghiasi2017,
benitez2018,omar2019,offidani2018,gmitra2017, khoo2017,zollner2020,luo2017,avsar2017,
island2019,safeer2019,ghiasi2019,benitez2020}. 
 Such graphene/TMD heterostructures possess a unique spin texture. In particular, the in-plane SOFs are of the Rashba type and point perpendicular to the electronic momentum. In the weak SOC regime, Rashba SOC caused by the stack inversion asymmetry leads to out-of-plane spin relaxation rates of $(\tper{})^{-1}=\Omega_\mathrm{R}^2\tp{}$, where $\Omega_\mathrm{R}$ is the Larmor frequency around the Rashba SOFs. 
 In contrast, the out-of-plane SOFs, which arise due to the broken sublattice symmetry in the TMD being imprinted on graphene, have opposite sign at the K and K' valleys (see Fig.~\ref{Figure1}a) to preserve time reversal symmetry \cite{gmitra2016}. These SOFs, commonly called valley-Zeeman SOFs, give rise to spin-valley locking. In this case, the intervalley scattering time (\tiv{}) is the characteristic time scale dominating the spin dynamics. Hence, in the weak SOC regime, the in-plane spin relaxation rate is given by $(\tpar{})^{-1}=\Omega_\mathrm{VZ}^2\tiv{}+(2\tper{})^{-1}$, where $\Omega_\mathrm{VZ}$ is the Larmor frequency around the valley-Zeeman SOFs.   
Since $\tiv{}$ is typically much longer than $\tau_p$ \cite{gorbachev2007}, $\tiv{}\Omega_\mathrm{VZ}$ becomes significantly bigger than $\tau_p{}\Omega_\mathrm{R}$ and, as a consequence, the spin transport is highly anisotropic  \cite{cummings2017,ghiasi2017,
benitez2018,omar2019,offidani2018}. 
Unlike in conventional materials, the spin-valley locking present in graphene/TMD heterostructures might enable the strong SOC regime if $\tvz{}$ would become comparable to $\tiv{}$, where \tvz{}$=2\pi/\Omega_\mathrm{VZ}$ is the in-plane spin precession period around the {out-of-plane} valley-Zeeman SOFs (Fig.~\ref{Figure1}b). Such condition may even be achieved in the diffusive regime and could allow for room temperature operations.

In this Letter, we report the achievement of the strong SOC regime in bilayer graphene (BLG)/WSe$_2$ heterostructures, leading to magnetic-field free spin precession induced by the valley-Zeeman SOC as shown in Figs.~\ref{Figure1}a and \ref{Figure1}b. Furthermore, by tuning the carrier density using a backgate voltage (\Vbg{}) and a drift current (\Idc{}), we control the spin polarization up to room temperature, making our device operate as a Datta-Das spin field-effect transistor (see Figs.~\ref{Figure1}c and \ref{Figure1}d). This hitherto unreported performance paves the way for the achievement of highly functional logic circuits \cite{dieny2020, sugahara2010}.

\begin{figure*}
	\centering
		\includegraphics[width=\textwidth]{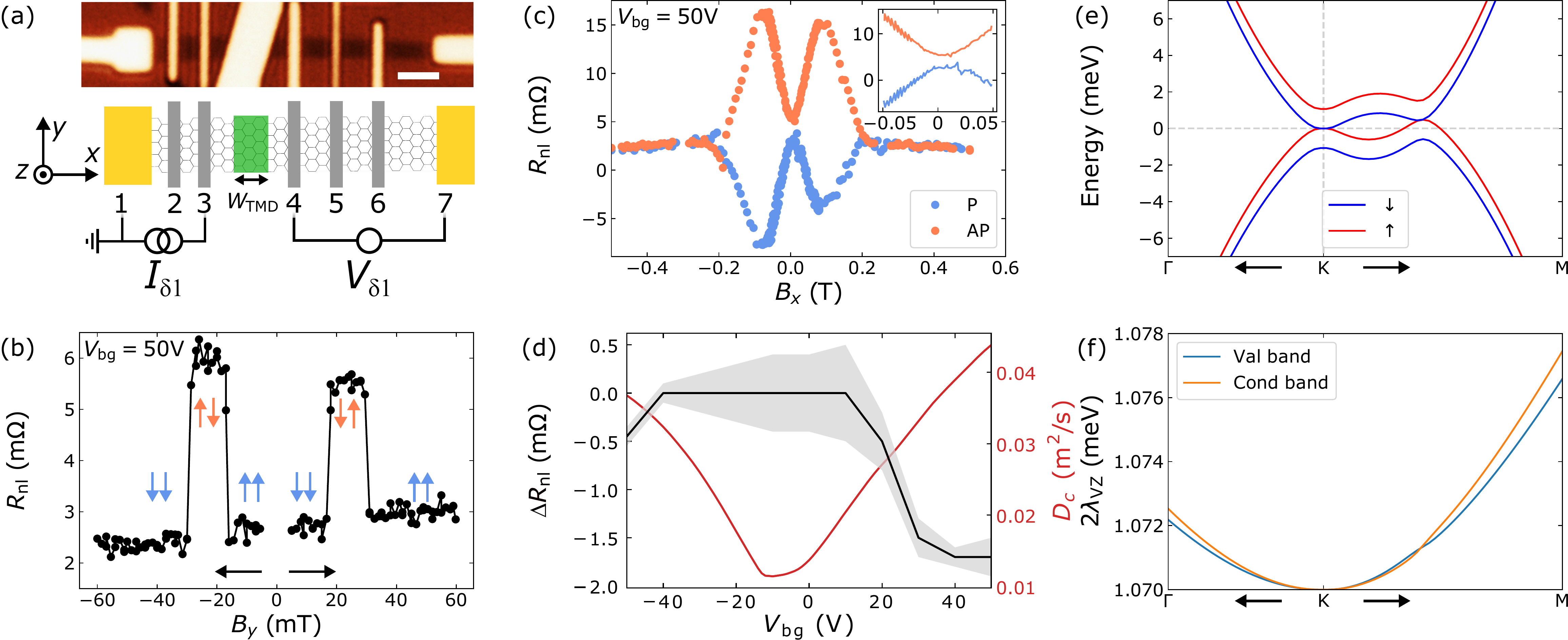}
	\caption{{Diffusive spin transport at 50 K.} (a) Optical image of the measured device. The BLG flake is the dark horizontal stripe and WSe$_2$ is the bright flake in the middle. The scale bar is 2~$\mu$m. The bottom panel shows a sketch of the device with the WSe$_2$-covered BLG region shown in green. The circuit corresponds to the standard nonlocal spin diffusion measurement configuration. Contacts 1 and 7 are not magnetic (Ti/Au) and 2 to 6 are spin-polarized TiO$_x$/Co contacts. (b) Nonlocal spin valve measurement across the WSe$_2$-covered BLG region as a function of the magnetic field applied along $y$ ($B_y$) for $\Vbg{} = 50$~V. The horizontal arrows represent the $B_y$-sweep direction and the vertical ones the magnetization of contacts 3 and 4. (c) Nonlocal spin precession measurements with the magnetic field applied along $x$ ($B_x$) in the parallel (P) and anti-parallel (AP) configurations for \Vbg{}$ = 50$~V. Inset: Low field detail. (d) Spin signal (\DRnl{}) and charge diffusivity (\Dc{}) as a function of \Vbg{}. 
(e) Spin-polarized band structure of BLG/WSe$_2$ at zero electric field. The red lines represent spin-up (along $+z$) and the blue ones spin-down (along $-z$). (f) Spin splitting ($2\lvz{}$) of the valence and conduction bands obtained from the band structure in panel e.}
	\label{Figure2}
\end{figure*}

To measure SOC-induced spin precession, we prepared 2-$\mu$m-wide (the heterostructure width is defined as $W_\mathrm{TMD}$ in Fig.~\ref{Figure2}a) BLG/WSe$_2$ lateral spin valves with spin-polarized TiO$_x$/Co contacts and Ti/Au reference electrodes (Fig.~\ref{Figure2}a). BLG was chosen to take advantage of its gate tunable diffusivity. To ensure an efficient SOC and achieve the strong SOC regime, we chose WSe$_2$, the TMD that imprints the largest valley-Zeeman SOC on graphene \cite{gmitra2016}, and annealed the van der Waals heterostructures at 430$^\circ$C. See Ref.~\cite{supinfo} for the fabrication details, {reproducibility}, the role of the annealing temperature, and the role of $W_\mathrm{TMD}$ on the measured signals.



The diffusive spin transport experiments are performed in the nonlocal geometry (circuit in Fig.~\ref{Figure2}a, see Ref.~\cite{supinfo} for measurement details). The $y$-spin accumulation ($\mu_{sy}$) induced by applying a current $I_{\delta1}$ through contact 3 diffuses across the channel and builds a voltage difference $V_{\delta1}=P_d\mu_{sy}/e$ between contacts 4 and 7. Here $P_d$ is the detector spin polarization and $e$ the electron charge. The nonlocal resistance ($\Rnl{}=V_{\delta1}/I_{\delta1}$) is measured as a function of a magnetic field applied along $y$ ($B_y$) in the conventional spin valve experiment. Figure~\ref{Figure2}b shows that, for \Vbg{}=$50$~V, \Rnl{} in the antiparallel magnetization state ($R_\mathrm{nl}^\mathrm{AP}$) is higher than in the parallel one ($R_\mathrm{nl}^\mathrm{P}$). The spin signal, which is defined as $\DRnl{}=(R_\mathrm{nl}^\mathrm{P}-R_\mathrm{nl}^\mathrm{AP})/2$, is thus negative. This observation could be a consequence of the sought in-plane spin precession induced by the valley-Zeeman coupling, although it could also be caused by the spin injector and detector having opposite spin polarizations \cite{kamalakar2016, xu2018}.

To confirm that $\mu_{sy}$ is reversed during transport as in Fig.~\ref{Figure1}b, we induce out-of-plane spin precession by measuring \Rnl{} as a function of a magnetic field applied along $x$ ($B_x$) (see Fig.~\ref{Figure2}c).  
In the parallel configuration, $R_\mathrm{nl}^\mathrm{P}$ has a local maximum at $B_x=0$, when the spins are not precessing. Then, $R_\mathrm{nl}^\mathrm{P}$ decreases until it reaches a minimum \emph{shoulder} ($B_x\approx\pm0.1 \, \mathrm{T}$) when the average precession angle at the detector is of 180$^\circ$. 
In this case, the spins injected along $y$ cross the TMD-covered region pointing along $z$, and reach the detector pointing along $-y$.  
At higher $B_x$, $R_\mathrm{nl}^\mathrm{P}$ increases until it merges with $R_\mathrm{nl}^\mathrm{AP}$ and \DRnl{} reaches zero as the spins dephase and the contact magnetizations are pulled towards $x$. In contrast, $R_\mathrm{nl}^\mathrm{AP}$ shows a minimum at $B_x=0$, where it is higher than $R_\mathrm{nl}^\mathrm{P}$. As $B_x$ increases, $R_\mathrm{nl}^\mathrm{AP}$ also increases leading to an enhancement of \DRnl{} with $B_x$ (Fig.~\ref{Figure2}c, inset). This result is in stark contrast with standard spin precession measurements (where \DRnl{} decreases at low $B$, until it reverses sign when the precessed angle is of 90$^\circ$ \cite{ghiasi2017,
benitez2018}) and is a direct consequence of $\mu_{sy}$ being reversed with respect to the out-of-plane spin accumulation. Finally, $R_\mathrm{nl}^\mathrm{AP}$ reaches a maximum when the precessed angle at the detector is of 180$^\circ$, before the contact magnetization pulling and spin dephasing decrease the spin signal until it vanishes for $B_x>0.2$~T. 
We observe that: (1) the magnitude of the in-plane spin signal ($B_x=0$) is significantly smaller than the out-of-plane one $(B_x\approx\pm0.1 \, \mathrm{T})$, in agreement with previous works in graphene/TMD heterostructures \cite{cummings2017,ghiasi2017,
benitez2018, omar2019, offidani2018}; (2) In contrast to the in-plane signal, the out-of-plane one is not reversed. This observation, together with the fact that out-of-plane spins are in the weak SOC regime \cite{supinfo}, indicates that the sign reversal is not caused by the opposite sign of the injector and detector spin polarizations.
Hence, $\mu_{sy}$ must be reversed during transport. Spin transport experiments performed at the pristine BLG region show conventional positive signal for all \Vbg{} values \cite{supinfo}, evidencing that the sign reversal occurs across the TMD-covered region. Since the in-plane spin signal is negative without an applied magnetic field, we conclude that our experiments are probing the strong SOC regime. Note that our result provides the most direct experimental evidence that spin precession occurs between scattering events in graphene/TMD heterostructures \cite{cummings2017,ghiasi2017,
benitez2018,omar2019, offidani2018,zihlmann2018,wakamura2018}.

In Fig.~\ref{Figure2}d (black curve), we plot the spin signal as a function of \Vbg{}. The data shows that the signal is negative for \Vbg{}$>\,20$ V and \Vbg{}$<\,-40$ V. For $-40\,\mathrm{V}<\Vbg{}<20\,\mathrm{V}$, \DRnl{} is below the noise level (see Ref.~\cite{supinfo} for the raw data). 
To understand the gate dependence, one must take into account the SOC in the BLG/WSe$_2$ heterostructure. As reported recently \cite{gmitra2017, khoo2017,zollner2020}, the SOC in BLG/TMD heterostructures can have a pronounced electric field dependence. To obtain the \Vbg{}-dependence of the SOC in our system, we have used the tight-binding Hamiltonian shown in Ref.~\cite{zollner2020} (see also Ref.~\cite{supinfo}). To explain the symmetric dependence of \DRnl{} vs \Vbg{} with respect to the charge neutrality point, we have assumed that both layers have the same potential, which means that the externally applied field compensates for the internal 0.267~V/nm induced by the WSe$_2$ on the BLG at the charge neutrality point \cite{gmitra2017}. The results from this band-structure calculation are displayed in Figs.~\ref{Figure2}e and \ref{Figure2}f and show perfect agreement with Ref.~\cite{gmitra2017}. As expected, the conduction and valence bands cross at the K point because of the layer-symmetric configuration. Looking at the spin splitting (2\lvz{}$=\hbar\Omega_\mathrm{VZ}$, where $\hbar$ is the reduced Plank constant) in Fig.~\ref{Figure2}f, we observe that it depends very weakly with the energy, indicating that the proximity SOC remains almost constant through the calculated energy range. This observation implies that the \Vbg{}-dependence of the SOC is unlikely to be the reason for the observed gate dependence. As shown in Fig.~\ref{Figure1}b and Ref.~\cite{supinfo}, if \tiv{} changes with \Vbg{} \cite{gorbachev2007}, it can tune the spin precession frequency but, since proximitized graphene shows weak antilocalization \cite{wang2016, yang2016, zihlmann2018, wakamura2018}, we could not measure weak localization in our device to extract \tiv{}.
In contrast, the charge diffusivity ($D_c$) of the BLG decreases significantly near the charge neutrality point (see red curve in Fig.~\ref{Figure2}d and Ref.~\cite{supinfo}). As shown by our spin transport calculations \cite{supinfo}, changes in $D_s$ (which we obtain assuming $D_s = D_c$ {\cite{maassen2011}}), can have a {crucial} influence on the spin signal in the strong SOC regime, making $D_s$ the most likely responsible for the measured \Vbg{} dependence. 
However, the electron-hole asymmetry in \DRnl{} indicates that other factors such as spin absorption by the WSe$_2$ \cite{yan2016,dankert2017} may also play a role. Note that we cannot discard a sign change of the signal below the noise level near the charge neutrality point.

 \begin{figure*}[tb]
	\centering
		\includegraphics[width=\textwidth]{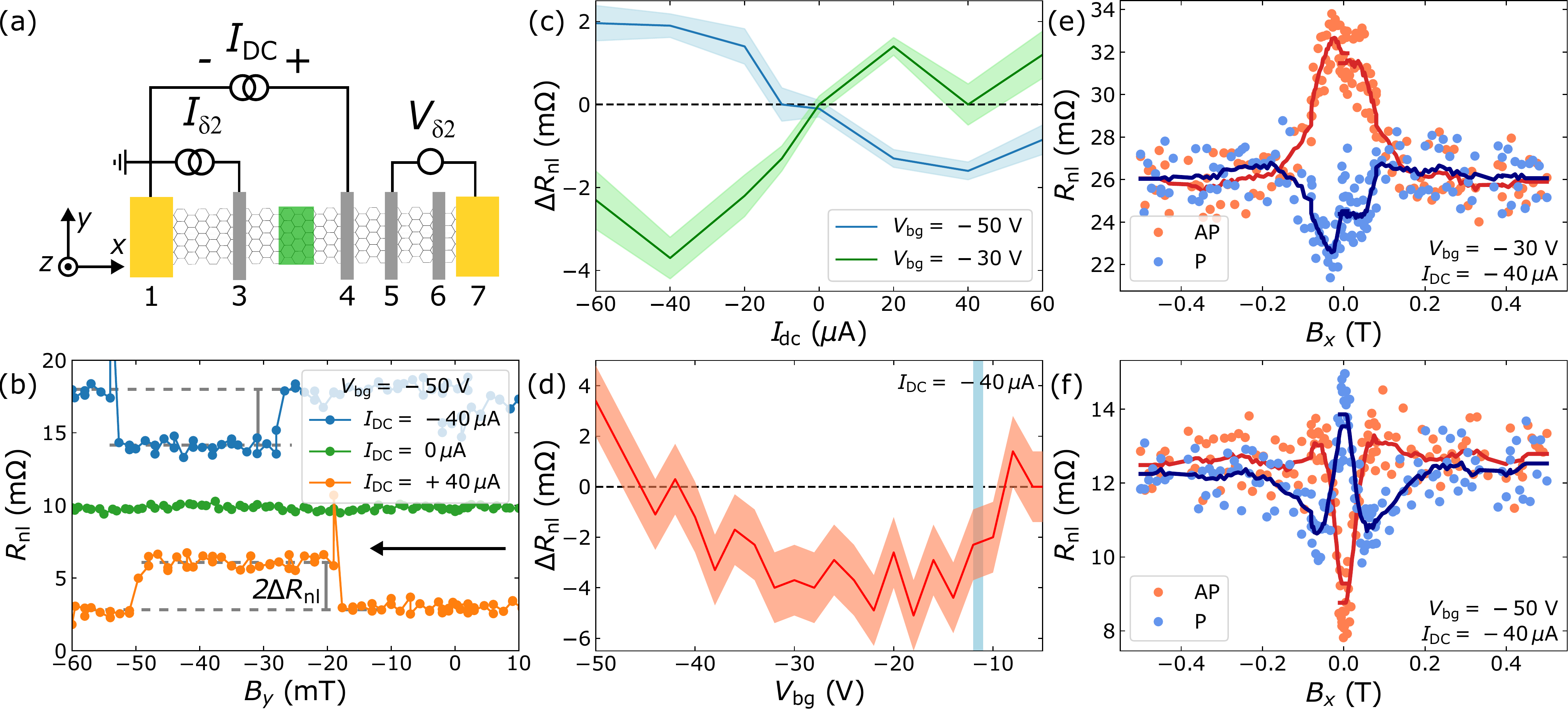}
	\caption{{Controlling spin transport with drift at 50 K.} (a) Sketch of the device with the spin drift 
	 measurement configuration. (b) Nonlocal spin valve measurements for \Vbg{} = $-50$~V at \Idc{} = $-40$, $0$, and $+40$~$\mu$A. The curves have been shifted for clarity. (c) \Idc{}-dependence of \DRnl{} at \Vbg{} = $-50$ and $-30$~V. (d) \DRnl{} vs \Vbg{} at \Idc{} = $-40$~$\mu$A. The vertical light blue line is the charge neutrality point of the WSe$_2$-covered BLG region. (e), (f) Nonlocal spin precession measurements with $B_x$ in the P and AP configurations for \Idc{} = $-40\,\mu$A and \Vbg{}$=\,-30$~V and $-50$~V, respectively. The lines are obtained by averaging over a window of eleven points. 
	 }
	\label{Figure3}
\end{figure*}

By tuning the spin dynamics in the strong SOC regime, it should be possible to control the \DRnl{} sign in a magnetic-field free device geometry. To confirm our hypothesis, we perform spin transport experiments under the effect of carrier drift in the geometry shown in Fig.~\ref{Figure3}a. The carrier drift is induced by \Idc{}, which is applied between contacts 4 and 1, and the spin current injected at contact 3 is detected as a nonlocal signal ($\Rnl{}=V_{\delta2}/I_{\delta2}$) between contacts 5 and 7. {Since $V_{\delta2}$ is coupled to $I_{\delta2}$, our measurement excludes the DC spin current injected by contact 4 \cite{supinfo}.}
The applied \Idc{} induces a drift velocity $v_d=\Idc{}/(W_\mathrm{BLG}ne)$, where $n$ is the carrier density in the channel and $W_\mathrm{BLG}$ is the BLG flake width. The induced $v_d$ changes the spin transport time across the BLG/WSe$_2$ \cite{YuFlatte, jozsa2008, ingla2016}, leading to a tuning of the spatial oscillation frequency of $\mu_{sy}$ \cite{supinfo}. In Fig.~\ref{Figure3}b, we present spin valve measurements at \Vbg{}$=\,-50$~V and \Idc{}$=\,-40$, $0$, and $ 40\,\mu$A. We observe that, in contrast with the results obtained from spin drift experiments in the pristine graphene region \cite{supinfo}, \DRnl{} reverses sign as we sweep \Idc{} from $-40$ to $40\,\mu$A, and becomes smaller than the noise level for \Idc{}$=\,0$. This result is the first demonstration of carrier drift-control of spin reversal in an all-electrical device. Such unprecedented observation is consistent with the spin transport model shown in Figs.~\ref{Figure1}a and \ref{Figure1}b {if the in-plane spin precession angle at \Idc{}=0 is a multiple of 90$^\circ$} (see Ref.~\cite{supinfo} {for more detailed calculations}). 
A comprehensive illustration of this behavior is shown in Fig.~\ref{Figure3}c, where we plot \DRnl{} vs \Idc{} as extracted from spin valve measurements performed at $\Vbg{}=\,-50$ and $-30$~V (see Ref.~\cite{supinfo} for the complete set of data). Importantly, we find that \DRnl{} reverses sign between the two \Vbg{} for all the \Idc{} values. To explain this sign reversal, we consider $\lvz{}$, \tiv{}, and $D_s$, that are the relevant parameters that could change with \Vbg{} (note that \tper{} and $n$, that changes $v_d$, cannot explain the observed sign reversal, see Ref.~\cite{supinfo} for details). 
We dismiss $\lvz{}$ because, according to our tight-binding calculations (Figs.~\ref{Figure2}e and \ref{Figure2}f), the valley-Zeeman SOC does not have a significant dependence with \Vbg{}. As mentioned above, \tiv{} may change with \Vbg{} \cite{gorbachev2007} and modify the effective spin precession frequency, as shown in Fig.~\ref{Figure1}b and \cite{supinfo}. Finally, we consider the change in $D_s$ from 0.01 to 0.03~m$^2$/s and observe that it has a strong influence on the $\mu_{sy}$ spatial frequency \cite{supinfo}. Even though both \tiv{} and $D_s$ could be responsible for the sign reversal of \DRnl{} with \Vbg{}, the extracted change in $D_s$ is large enough to explain a sign reversal keeping \tiv{} constant.  

Our observation of a sign reversal in \DRnl{} with \Vbg{} at fixed \Idc{} is very promising for Datta-Das spin field-effect transistor operations which work in the diffusive regime, as sketched in Figs.~\ref{Figure1}c and \ref{Figure1}d. In Fig.~\ref{Figure3}d, we plot \DRnl{} vs \Vbg{} at \Idc{}$=-40\,\mu$A. We find that \DRnl{} becomes positive for \Vbg{}$<\,-45$~V and at \Vbg{}$=\,-8$~V. 

\begin{figure}[tb!]
\centering
		\includegraphics[width=0.4\textwidth]{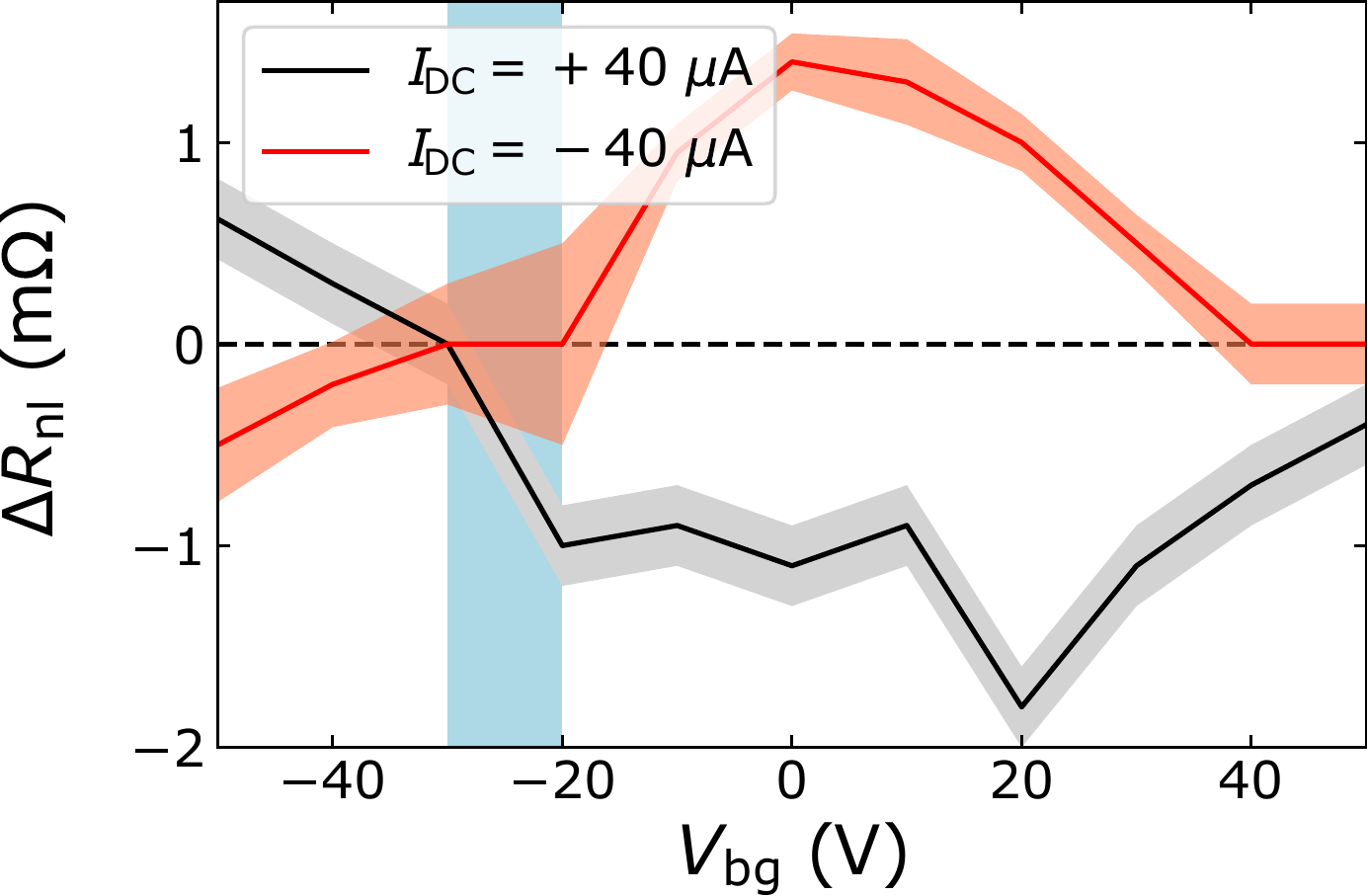}
	\caption{{Room temperature electrical control of spin transport.} 
	\Vbg{}-dependence of \DRnl{} for $\Idc{}=\,\pm40\,\mu$A. The blue area represents the charge neutrality point of the WSe$_2$-covered BLG region. As at 50 K, \DRnl{} reverses sign upon changing the sign of \Idc{}. 	
}
	\label{Figure4}
\end{figure}
Finally, we measure spin precession around $B_x$ to confirm that the out-of-plane spin signal has not changed sign and the previous results are indeed caused by in-plane spin precession. The results are shown in Figs.~\ref{Figure3}e and \ref{Figure3}f for \Vbg{}$=\,-30$ and $-50$~V, respectively. For \Vbg{}$=\,-30$~V, the spin precession data looks similar to the one in Fig.~\ref{Figure2}c with the difference that the in-plane $B_x=0$ signal in Fig.~\ref{Figure3}e is comparable to the maximum signal at the shoulders. As a consequence, the shoulders are less clear than in Fig.~\ref{Figure2}c. In contrast, the \Rnl{} vs $B_x$ data at \Vbg{}$=\,-50$~V shows a conventional spin precession shape where the in-plane spin signal is positive and larger than \DRnl{} at the shoulders, {more} similar to isotropic systems \cite{ghiasi2017,benitez2018}. See Ref~\cite{supinfo} for the evolution of the spin precession data with \Idc{}.

To confirm that the measured effect is suitable for applications, we perform spin valve experiments at 300~K as a function of \Vbg{} (see Ref.~\cite{supinfo} for the raw data). 
The \DRnl{} values are plotted in Fig.~\ref{Figure4} for \Idc{}$=\,\pm40\,\mu$A. These results are very similar to those at 50~K, demonstrating that our device is in the strong SOC regime up to room temperature and the spin orientation can be controlled using both \Idc{} and \Vbg{}. Similar results obtained in a second sample are shown in Ref.~\cite{supinfo}.

To conclude, we demonstrate the valley-Zeeman SOC induced magnetic-field free control of spin precession in a BLG/WSe$_2$ van der Waals heterostructure at the strong SOC regime. By tuning the carrier density using \Vbg{} and \Idc{}, we control the spin polarization up to room temperature, making our device operate as a spin field-effect transistor. This achievement has prospect for future spin-based logic applications such as nonvolatile and reconfigurable logic \cite{sugahara2010} and as a complement to the existing spin-logic proposals \cite{dery2007,behin2010,manipatruni2019,pham2020}. 


\section{Acknowledgments}
We acknowledge R.~Llopis and R.~Gay for technical assistance and C.~K.~Safeer, N.~Ontoso, and K. Zollner for discussions. This work is supported by the Spanish MICINN under the Maria de Maeztu Units of Excellence Programme (MDM-2016-0618) under Project RTI2018-094861-B-100 and by the the European Union H2020 under the Marie Slodowska Curie Actions (0766025-QuESTech). J. F. acknowledges Deutsche Forschungsgemeinschaft  (DFG,  German  Research  Foundation)  SFB1277
(Project-ID 314695032), SPP 2244, and the European Unions Horizon 2020 research and innovation program under Grant No.  785219. J.I.-A. acknowledges postdoctoral fellowship support from the “Juan de la Cierva - Formación” program by the Spanish MICINN (Grant No. FJC2018-038688-I).


\bibliography{bibliography}

\newpage
\textcolor{white}{Dummy text}
\newpage

\widetext
\begin{center}
\textbf{\large Supplemental Material of "Electrical control of valley-Zeeman spin-orbit coupling-induced spin precession at room temperature"}
\end{center}
\setcounter{equation}{0}
\setcounter{figure}{0}
\setcounter{table}{0}
\setcounter{page}{1}
\makeatletter
\renewcommand{\theequation}{S\arabic{equation}}
\renewcommand{\thefigure}{S\arabic{figure}}
\newpage
\tableofcontents

\clearpage
\section{Device fabrication}
\begin{figure}[h]
	\centering
		\includegraphics[width=0.8\textwidth]{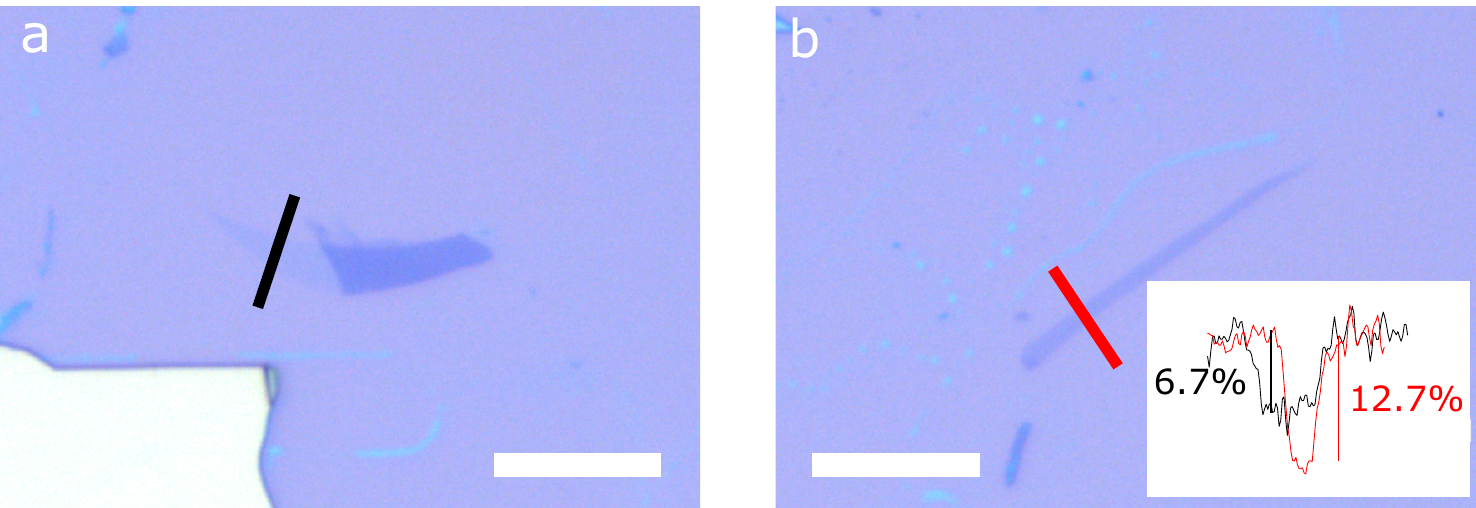}
	\caption{{Optical contrast of BLG.} (a) Optical microscope image of a monolayer graphene flake. (b) Optical microscope image of the BLG flake used to prepare the device shown in the main text. Inset: Intensity profiles taken along the black and red lines with the respective colors. The optical contrast of the monolayer is 6.7\% and the bilayer 12.7\%. The scale bars are 10~$\mu$m.}
	\label{FigureOpticalContrast}
\end{figure}
The BLG flake was obtained by cleaving a highly oriented pyrolytic graphite crystal (provided by HQ graphene) on a Si substrate with 300 nm of thermal oxide using Nitto SPV 224P tape.To realize spin transport in a channel with tuneable diffusivity, we  select bilayer graphene (BLG) as the optimal channel material. To determine the number of layers of the exfoliated flakes, we used optical contrast. In particular, the optical contrast of BLG is twice the one of monolayer graphene (see Fig.~\ref{FigureOpticalContrast}).
The WSe$_2$ was exfoliated from a crystal (by HQ graphene) on a PDMS stamp (gelpack 4). The 27-nm-thick WSe$_2$ flake was transferred on top of the graphene flake using the viscoelastic stamping technique \cite{castellanos2014} and the resulting heterostructure was annealed for 1 h at 430 $^\circ$C under high vacuum conditions. Next, we prepared the Ti(5~nm)/Au(35~nm) contacts using standard e-beam lithography, e-beam deposition of Ti and thermal deposition of Au. Finally, after being defined by e-beam lithography, the TiO$_x$/Co contacts were prepared by depositing 3~\angstrom{} of Ti and, after oxidation in ambient conditions, 35~nm of Co were deposited  by e-beam evaporation. Finally, 5~nm of Au were deposited to cap the Co layer, and the whole structure was covered by an insulating hexagonal boron nitride flake. Figure~\ref{FigureDeviceImage}a shows an optical micrograph image of the device before the electronic measurements. 
 \section{Electronic measurements}\label{SectionElMeas}
 The spin (charge) transport measurements were performed using the DC reversal (delta) mode of a Keithley~6221 current source and a Keithley~2182 nanovoltmeter with a delay of 20~ms and an excitation current of $I_{\delta1(2)}=80~\mu$A (1~$\mu$A for charge transport). The DC reversal technique allows us to measure small signals while removing backgrounds, enabling for the measurement of the nonlocal spin signal $V_{\delta 1}$ in the circuit of Fig.~2a of the main manuscript. Furthermore, the DC reversal technique also allows us to remove the DC spins injected by contact 4 in the drift current geometry shown in Fig.~3a of the main manuscript. Additionally, the carrier density of the graphene channel is tuned by applying a \Vbg{} to the doped Si substrate (see Section~S2). Both \Idc{} and \Vbg{} are applied using a Keithley~2636B system source meter. The magnetic field is applied using a superconducting solenoid and  a rotator to orient magnetic fields along the $x$ and $y$ directions defined in Fig.~2a of the main manuscript.
\section{Analysis of the \Vbg{} sweeps}\label{SectionGateSweep}
\begin{figure}[h]
	\centering
		\includegraphics[width=0.4\textwidth]{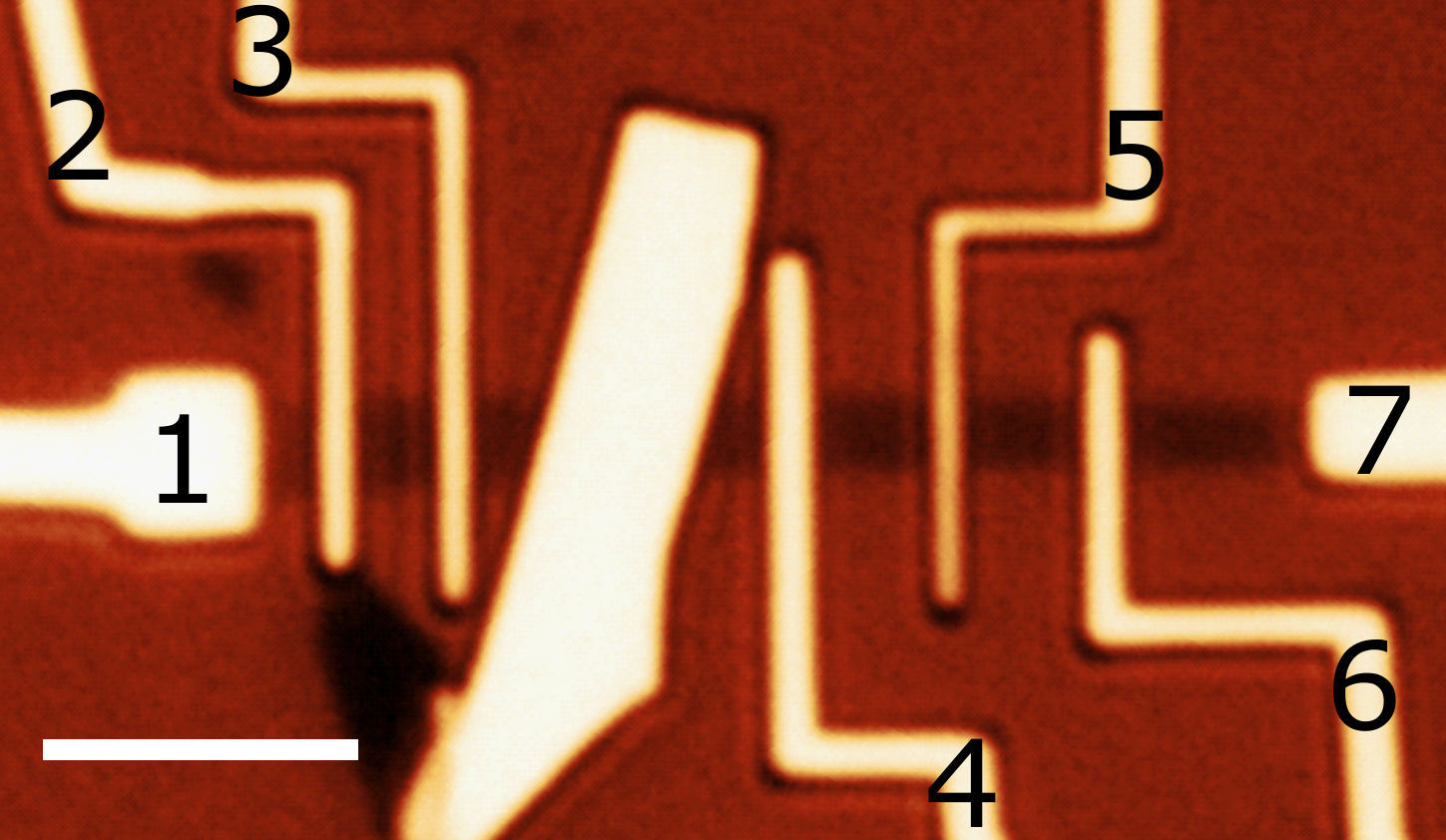}
	\caption{{Optical microscope image of the device with the corresponding contact numbering.} The scale bar is 4~$\mu$m.}
	\label{FigureDeviceImage}
\end{figure}
\subsection{Measurement of the square resistance of the pristine and WSe$_2$-covered BLG regions}
\begin{figure}[h]
	\centering
		\includegraphics[width=0.75\textwidth]{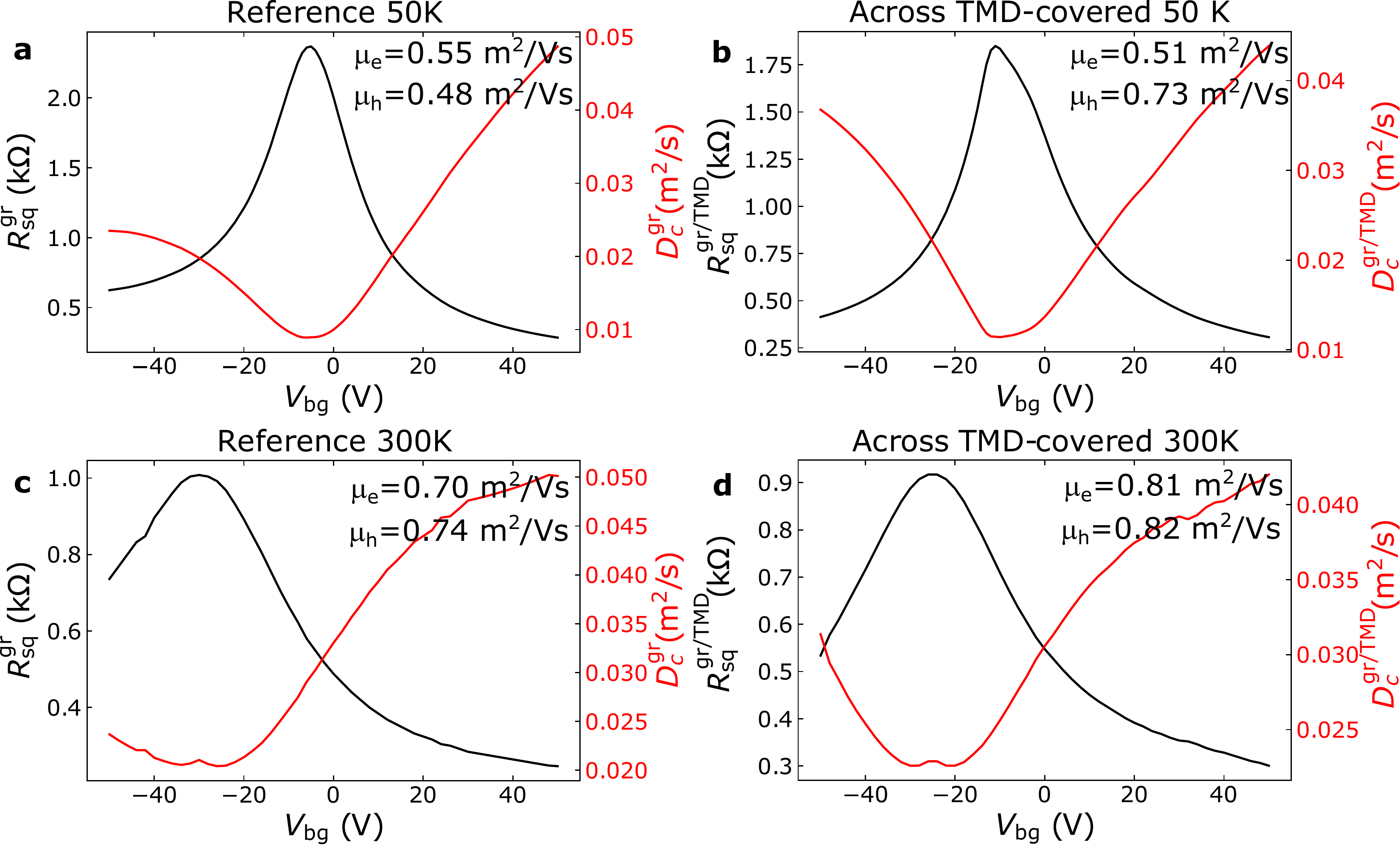}
	\caption{{Charge transport characterization.} (a) Square resistance \Rgr{} and charge diffusivity $D_c^\mathrm{gr}$ of the pristine BLG region as a function of \Vbg{} at 50~K. (b) \RgrTMD{} and $D_c^\mathrm{gr/TMD}$ obtained across the WSe$_2$-covered region vs \Vbg{} at 50~K. (c) \Rgr{} and $D_c^\mathrm{gr}$ vs \Vbg{} at 300~K. (d) \RgrTMD{} and $D_c^\mathrm{gr/TMD}$ vs \Vbg{} at 300~K.}
	\label{FigureRsq}
\end{figure}

To determine the charge transport properties of the different regions of the device shown in the main manuscript and Fig.~\ref{FigureDeviceImage}, we measured the channel's square resistance \Rsq{} as a function of the backgate voltage (\Vbg{}), that is applied to the doped Si substrate \cite{novoselov2004}. The \Vbg{} controls the carrier density ($n$) in the graphene channel via the field effect,

\begin{equation}
n=\frac{\epsilon_0\epsilon_r}{et_\mathrm{SiO_2}}(V_\mathrm{bg}-V_\mathrm{cnp}),
\label{EquationS1}
\end{equation}
where $\epsilon_{0}$ is the vacuum dielectric permittivity, $\epsilon_{r}=3.9$ is the dielectric constant of SiO$_2$, $e$ the electron charge, $t_\mathrm{SiO_2}=300$~nm is the thickness of the SiO$_2$ dielectric, and $V_\mathrm{cnp}$ the value of \Vbg{} at which the graphene reaches the charge neutrality point (CNP).
In Fig.~\ref{FigureRsq}a, we show \Rsq{} vs \Vbg{} at the pristine graphene region (\Rgr{}) at 50~K obtained by measuring the voltage drop between contacts 5 and 4 ($V_{54}$) while applying a current between contacts 7 and 1 ($I_{71}=1\,\mu$A). \Rgr{} is determined using 
\begin{equation}
R_\mathrm{sq}^\mathrm{gr}=(V_{54}/I_{71})(W_\mathrm{gr}^{54}/L_{54}),
\label{EquationS2}
\end{equation}
where $W_\mathrm{gr}^{54}=0.80\,\mu$m is the average sample width between contacts 5 and 4 and $L_{54}=2.00\,\mu$m is the spacing between contacts 5 and 6 (Table~\ref{TableS1}). From Fig.~\ref{FigureRsq}a, one can observe that \Rsq{} shows a clear peak for \Vbg{}$=V_\mathrm{cnp}=\,-$5~V, which corresponds to the CNP of this region.

Next, we study the \Vbg{}-dependence of the \Rsq{} of the WSe$_2$-covered graphene region (\RgrTMD{}). For this purpose, we measure the voltage between contacts 4 and 3 ($V_{43}$) while applying a DC current between contacts 7 and 1 ($I_{71}=1\,\mu$A). We obtain \Rsq{} using Equation~\ref{EquationS2} and changing $V_{54}$ and $L_{54}$ with $V_{43}$ and $L_{43}=4.30\,\mu$m. Finally, $W_\mathrm{gr}^{43}=0.85\,\mu$m is the average graphene flake width between contacts 3 and 4. \RgrTMD{} vs \Vbg{} is shown in Fig.~\ref{FigureRsq}b and shows a peak at \Vbg{}$=V_\mathrm{cnp}=\,-$11~V and an upturn at $\Vbg{}=\,-5$~V, which we attribute to the pristine graphene region between contacts 4 and 5. At 300~K, as shown in Figs.~\ref{FigureRsq}c and \ref{FigureRsq}d, we obtain similar results with lower \Rsq{} values near the CNP. We attribute this observation to the thermal energy broadening, which leads to an increased residual carrier density (and hence conductivity). From these measurements, we cannot rule out the opening of a small bandgap in BLG induced by a combination of the weak electric field applied by the \Vbg{}\cite{mccann2013,zhang2009} and the field induced by the WSe$_2$\cite{gmitra2017,khoo2017}. However, the similar \Rsq{} at the charge neutrality point obtained at the WSe$_2$-covered and pristine BLG regions shows that the TMD is not opening a significant bandgap. Note that $V_\mathrm{CNP}$ also changes with temperature. 
\begin{table}[t]
    \caption{Geometrical device parameters.}
        \begin{ruledtabular}
        \begin{tabular}{c c c c c c}
            \renewcommand{\arraystretch}{2}
            $W_\mathrm{gr}^{43}$ & $L_{43}$&$W_\mathrm{gr}^{54}$&$L_{54}$& $W_\mathrm{TMD}$\\
			($\mu$m)	&($\mu$m)&($\mu$m)	&($\mu$m)&($\mu$m)\\
            \hline
             0.85 & 4.30 &0.80&2.00&2.00\\
    \end{tabular}
    \end{ruledtabular}
    \label{TableS1}
\end{table}
\subsection{Determination of the charge diffusivity}
Because of the weak electron-electron interactions in graphene \cite{bandurin2016}, in samples with moderate mobility, the charge ($D_c$) and spin diffusivity ($D_s$) have been shown to be equal \cite{maassen2011}. Hence, it is useful to obtain $D_c$ from the \Vbg{} sweeps. For this purpose, we use the Einstein relation $D_c=(e^2R_\mathrm{sq}\nu(E_F))^{-1}$, where $\nu(E_F)$ is the density of states at the Fermi level. Using the density of states of (ungapped) bilayer graphene, we obtain the following expression: 
\begin{equation}
D_c=\frac{\pi\hbar^2 v_{f0}^2}{R_\mathrm{sq}e^2\sqrt{\gamma_1^2+4\pi\hbar^2v_{f0}^2|n|}},
\label{EinsteinRelation}
\end{equation}
 
where $v_{f0}=1\times10^6$~m/s is the Fermi velocity of graphene, $\gamma_1\sim0.4$~eV is the interlayer coupling parameter between pairs of orbitals on the dimmer sites in BLG \cite{mccann2013}, and $\hbar$ is the reduced Planck constant. Using Equation~\ref{EinsteinRelation}, the measured \Rsq{}, and $n$ (obtained using Equation~\ref{EquationS1}) we obtain $D_\mathrm{c}^\mathrm{gr}$ for the pristine graphene region and $D_\mathrm{c}^\mathrm{gr/TMD}$ for the WSe$_2$-covered one. These results are shown as red lines in Figs.~\ref{FigureRsq}a-d. Note that, since we have performed the calculation assuming that the region between contacts 3 and 4 is all covered, our calculation is not very accurate near the CNP for Figs.~\ref{FigureRsq}b and \ref{FigureRsq}d.
Finally, to determine the charge transport quality of our device, we calculated the field-effect electron (hole) mobility ($\mu_{e(h)}$) using $\Rsq{}^{-1}=ne\mu_{e(h)}$, at $\Vbg{}-V_\mathrm{cnp}=+(-)15$~V. The results are labeled in Fig.~\ref{FigureRsq}. We observe that the obtained mobilities are of about 5000~cm$^2/$(Vs), as expected for graphene on SiO$_2$ devices. At 300~K, the mobilities are slightly higher than at 50~K, which we attribute to the broadening that leads to an underestimation of the carrier density close to the CNP. This is supported by the trend of $D_\mathrm{c}^\mathrm{gr}$ and $D_\mathrm{c}^\mathrm{gr/TMD}$, that is higher at 300~K than at 50~K for all the \Vbg{} range except for $\Vbg{}\gtrsim40$~V, where  $D_\mathrm{c}^\mathrm{gr}$ and $D_\mathrm{c}^\mathrm{gr/TMD}$ at 300~K start to saturate.
\subsection{Determination of the momentum scattering time and the Rashba SOC period}
To confirm that the out-of-plane spins are in the weak spin-orbit coupling (SOC) regime, we determine the momentum scattering time ($\tau_p$) from the charge transport measurements and compare it with the DFT values for the Rashba SOC precession frequency ($\Omega_\mathrm{R}$).
To determine the momentum scattering time \tp{} in BLG, we use $D_c=v_f^2\tp{}/2$, where $v_f$ is the Fermi velocity, that can be calculated using
\begin{equation}
v_f=\frac{1}{\hbar}\frac{dE(k)}{dk}=\frac{2\hbar v_{f0}^2|k|}{\gamma_1}=\sqrt{2}v_{f0}\sqrt{\sqrt{1+\frac{4|n|}{\alpha\gamma_1^2}}-1},
\end{equation}
where $k$ is the reciprocal lattice vector with respect to the K and K' points and $\alpha=(\pi\hbar^2v_{f0}^2)^{-1}$. We replaced $|k|$ by $|n|$ using $E=\hbar^2v_{f0}^2|k|^2/\gamma_1$ and $n=\int_0^{E_F}\nu(E)dE$.
 
At 50~K and $\Vbg{}=\,-50$~V, since the CNP of the WSe$_2$-covered BLG region is $V_\mathrm{cnp}=\,-11$~V, the carrier density $n\approx4.4\times10^{16}$~m$^{-2}$. Hence, $v_f\approx1.06\times10^{6}$~m/s. Since $D_c$($\Vbg{}=50\,\mathrm{V}$)$\approx4.4\times10^{-2}$~m$^2$/s, $\tau_p=2D_c/v_f^2\approx78$~fs. 

From ab initio calculations, $\lambda_\mathrm{R}^\mathrm{grWSe_2}=0.56$~meV on monolayer graphene (Ref.~\onlinecite{gmitra2016}). Since the calculation is performed at high carrier densities, we can safely assume no layer polarization is present \cite{gmitra2017, khoo2017}. Hence, to obtain the average Rashba SOC, we have to divide the monolayer graphene value by two $\lambda_\mathrm{R}^\mathrm{BLGWSe_2}=\lambda_\mathrm{R}^\mathrm{grWSe_2}/2=0.28$~meV yielding a $\Omega_\mathrm{R}=2\lambda_\mathrm{R}^\mathrm{BLGWSe_2}/\hbar=0.85\times10^{12}$~rad/s and $\tau_\mathrm{R}=2\pi/\Omega_\mathrm{R}=7.4$~ps. Hence, $\tau_p/\tau_\mathrm{R}\approx1\times10^{-2}$, which is much smaller than 1, confirming that the Rashba SOC is in the weak SOC regime.
 
Note that the proximity-induced spin-orbit parameters can depend on the rotation between the graphene and TMD layers \cite{li2019,david2019}. However, the maximum value of $\lambda_\mathrm{R}^\mathrm{WSe_2}$ reported in Ref.~\onlinecite{li2019} is $\approx0.88$~meV (note that $\lambda_\mathrm{R}^\mathrm{WSe_2}$ has been normalized here by a factor of two to match the definitions with the other references followed here), leading to $\tau_p/\tau_\mathrm{R}^{min}\approx3\times10^{-2}$. We conclude that our results would apply for any rotation value if the interlayer distance corresponds to the DFT-calculated one.

\section{Spin transport at the pristine graphene region at 50~K}\label{SectionOuterHanle}

 \begin{figure}[tbh!]
	\centering
		\includegraphics[width=\textwidth]{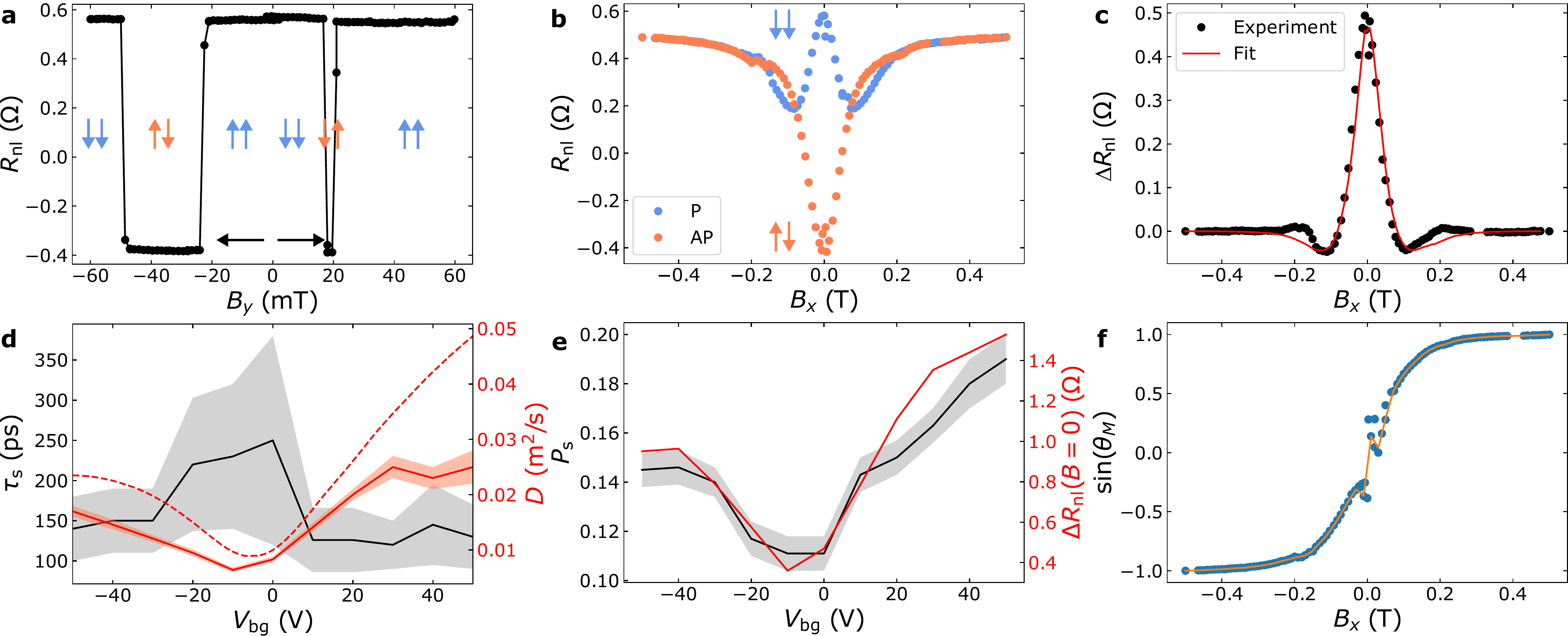}
	\caption{{Spin transport at the pristine graphene region at 50~K.} (a) Nonlocal spin valve measurement, \Rnl{}$=V_{57}/I_{41}$ as a function of $B_y$. (b) Nonlocal spin precession around $B_x$ in the parallel (P) and antiparallel (AP) magnetic contact configurations ($R_\mathrm{nl}^\mathrm{P}$, and $R_\mathrm{nl}^\mathrm{AP}$, respectively) at \Vbg{}$=0$~V. (c) $\DRnl{}=(R_\mathrm{nl}^\mathrm{P}-R_\mathrm{nl}^\mathrm{AP})/2$ as a function of $B_x$ obtained from the data in panel b (black solid circles) with its fit to the Bloch equations (red line). (d) Spin lifetime ($\tau_s$, black line), spin ($D_s$, red solid line) and charge ($D_c$, red dashed line) diffusivity. (e) Contact spin polarization ($P_s=\sqrt{P_iP_d}$) together with $\DRnl{}(B=0)$. (f) $\sin(\theta_M)$ obtained from the data in panel b (blue solid circles). The orange line is a guide to the eye.}
	\label{FigureRefHanle}
\end{figure}
We performed measurements in the nonlocal configuration in the pristine graphene region to determine the spin transport properties of our system at 50~K.
In particular, we applied a current between contacts 4 and 1 ($I_{41}$) and measure the nonlocal voltage between contacts 5 and 7 ($V_{57}$). The spin valve and spin precession measurements are shown in Figs.~\ref{FigureRefHanle}a and \ref{FigureRefHanle}b, respectively. From Fig.~\ref{FigureRefHanle}a, we observe that the spin signal is around 0.5~$\Omega$ and contacts 4 and 5 switch at very similar $B_y$ values for positive $B_y$, giving rise to a very narrow antiparallel state. Note that, in contrast with the spin transport measurements performed across the WSe$_2$-covered BLG region (see Fig.~2b of the main manuscript), \Rnl{} is higher for the parallel (P) than the antiparallel (AP) state, as expected for standard nonlocal measurements.
The next step is to apply $B_x$ to induce spin precession. We performed this experiment by taking the sweeps for negative and positive $B_x$ to control the magnetization state separately. Figure~\ref{FigureRefHanle}b shows the nonlocal resistance in the P ($R_\mathrm{nl}^\mathrm{P}$) and AP ($R_\mathrm{nl}^\mathrm{AP}$) magnetization configurations as indicated by the arrows. At low $B_x$, the spins injected along $y$ precess around the magnetic field leading to the crossing between the P and AP curves at $B_x\approx\pm65$~mT. As $B_x$ increases, the contact magnetizations get pulled towards $x$. The spins that are now polarized along $x$ do not precess, giving rise to the saturation of both $R_\mathrm{nl}^\mathrm{P}$ and $R_\mathrm{nl}^\mathrm{AP}$ at the value of $\Rnl{}=R_\mathrm{nl}^\mathrm{P}(B_x=0)$. Taking the contact pulling into account, the measured signal is 
\begin{equation}
R_\mathrm{nl}^\mathrm{P(AP)}=\pm\cos^2(\theta_M)f_{prec}+\sin^2(\theta_M)f_{prec}(B=0).
\end{equation}
Here, $f_{prec}$ is the spin precession signal, $\theta_M$ is the contact magnetization angle with respect to the easy axis.  

We first isolate the spin precession signal by plotting $\DRnl{}=(R_\mathrm{nl}^\mathrm{P}-R_\mathrm{nl}^\mathrm{AP})/2$ in Fig.~\ref{FigureRefHanle}c. To obtain the spin transport parameters of this region, we fit the spin precession data to $\DRnl{}=\cos^2(\theta_M)f_{prec}$. Here we use the solution of the Bloch equations for a homogeneous system taking into account the spin backflow in the spin injector and detector contacts 4 and 5 as $f_{prec}(P,\tau_s^\mathrm{gr},D_s^\mathrm{gr},R_{ci}, R_{cd},L_{54},W_s^{54})$, which is a function of $P=\sqrt{P_iP_d}$, where $P_{i(d)}$ is the spin injection (detection) efficiency, $\tau_s^\mathrm{gr}$ the spin lifetime, $D_s^\mathrm{gr}$ the diffusivity of the pristine graphene region, $R_{ci(d)}$ the contact resistance of the spin injector (detector) (Refs.~\onlinecite{maassen2012, idzuchi2015}). $\lambda_s^\mathrm{gr}=\sqrt{\tau_s^\mathrm{gr}D_s^\mathrm{gr}}$ is the spin relaxation length. 

To perform the fit, one needs to know $\theta_M$. We obtain it by using Equation~\ref{EquationContPull}:
\begin{equation}
sin(\theta_M)=\mathrm{sgn}(B_x)\sqrt{\frac{R_\mathrm{nl}^\mathrm{avg}-\min(R_\mathrm{nl}^\mathrm{avg})}{\max(R_\mathrm{nl}^\mathrm{avg})-\min(R_\mathrm{nl}^\mathrm{avg})}}
\label{EquationContPull}
\end{equation} 
where $R_\mathrm{nl}^\mathrm{avg}=(R_\mathrm{nl}^\mathrm{P}+R_\mathrm{nl}^\mathrm{AP})/2$, $\mathrm{sgn}(B_x)$ is the sign of the magnetic field, and $\max(R_\mathrm{nl}^\mathrm{avg})$ and $\min(R_\mathrm{nl}^\mathrm{avg})$ are the maximum and minimum values of $R_\mathrm{nl}^\mathrm{avg}$, respectively. The result of this operation is shown in Fig.~\ref{FigureRefHanle}f and is used to fit the data, shown as the red line of Fig.~\ref{FigureRefHanle}c. We have performed this operation for $\Vbg{}=\,-50$ to $50$~V. In Fig.~\ref{FigureRefHanle}d, we plot the extracted $\tau_s^\mathrm{gr}$ (black line) with its error range (grey area) and $D_s^\mathrm{gr}$ (red line) and its error range (light orange area). The dashed line corresponds to $D_c^\mathrm{gr}$ (Fig.~\ref{FigureRsq}a), which is in reasonable agreement with $D_s^\mathrm{gr}$ as expected. The apparent saturation of $D_s^\mathrm{gr}$ at positive \Vbg{} is likely caused by the contact pulling, that determines the data at high $B_x$. We also note that $D_s^\mathrm{gr}$ and $\tau_s^\mathrm{gr}$ show opposite trend with \Vbg{}. Hence, $\lambda_s^\mathrm{gr}$ shows little change with \Vbg{}. As a consequence, the contact spin polarization ($P_s=\sqrt{P_iP_d}$) shown in Fig.~\ref{FigureRefHanle}e follows closely the spin signal magnitude ($\DRnl{}(B=0)$). Finally, we note that the spin signal is positive for all the \Vbg{} values, ranging from 0.4 up to 1.5~$\Omega$. 
This result is also consistent with the WSe$_2$-covered region being responsible for the sign reversal.

\section{Spin transport across the WSe$_2$-covered graphene region at 50 K}
 \begin{figure}[tb!]
	\centering
		\includegraphics[width=0.7\textwidth]{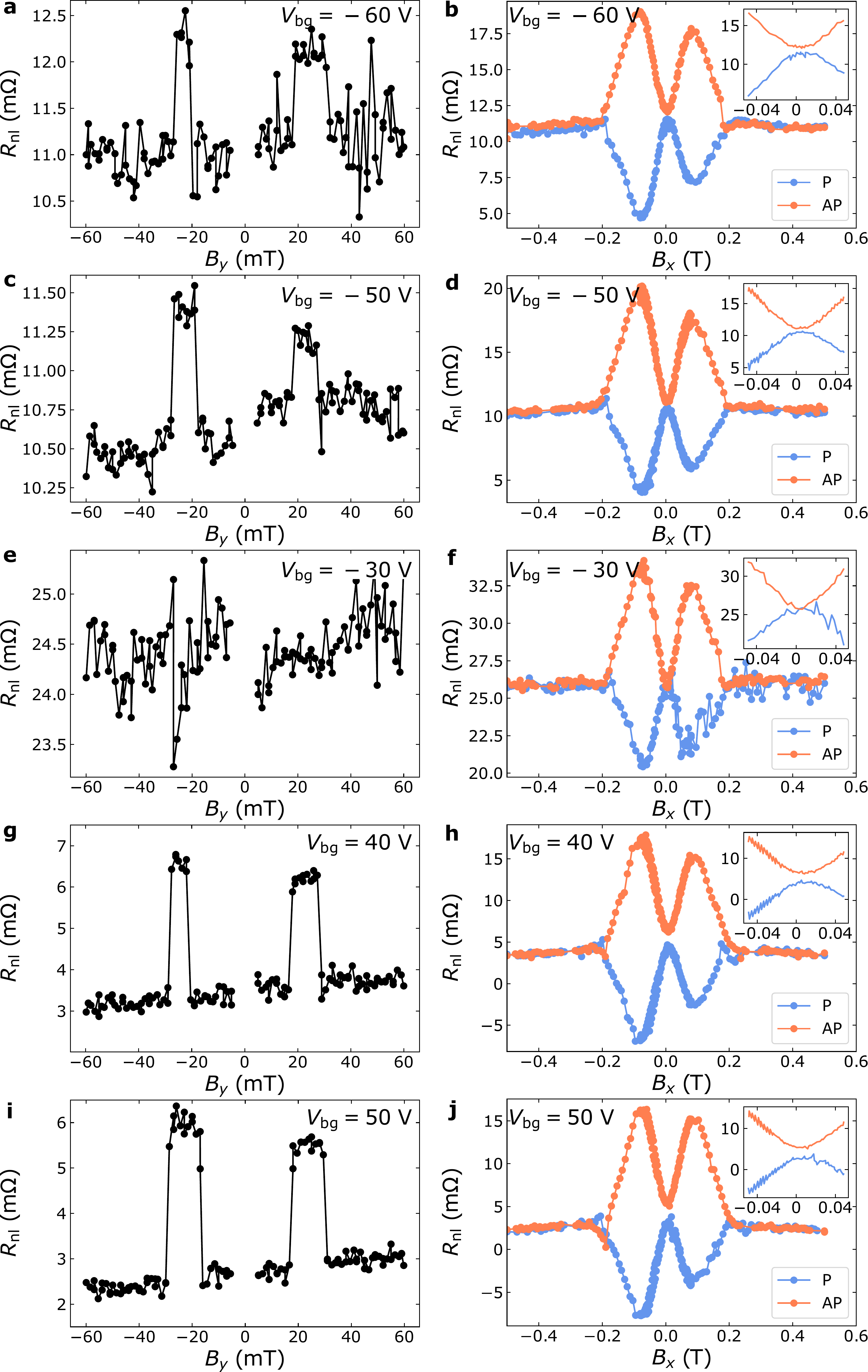}
	\caption{{Spin transport across the WSe$_2$-covered BLG region at 50~K.} (a), (c), (e), (g) and (i) Nonlocal spin valve measurements and (b), (d), (f), (h) and (j) nonlocal spin precession measurements performed at  \Vbg{}$=-60,\,-50,\,-30,\,40,$ and $50$~V. The insets are a zoom of the low $B_x$ range.}
	\label{FigureDifHanle}
\end{figure}
To study spin transport in the WSe$_2$-covered region at 50~K, we measured the nonlocal resistance by applying a current $I_{\delta1}$ between contacts 3 and 1 and measuring the voltage $V_{\delta1}$ between contacts 4 and 7. The nonlocal spin valve measurements are performed by taking \Rnl{}$=V_{\delta1}/I_{\delta1}$ as a function of the magnetic field applied along the $y$ direction ($B_y$). As we sweep $B_y$, the contact magnetizations are controlled independently due to the different coercivities of the contacts. In Fig.~\ref{FigureDifHanle}a, we can observe two abrupt changes of \Rnl{} for positive and negative $B_y$, which correspond to switches of the magnetization of contacts 4 and 3.
The $R_\mathrm{nl}^\mathrm{AP}$ is higher than $R_\mathrm{nl}^\mathrm{P}$.  
In Figs.~\ref{FigureDifHanle}a, c, e, g, and i, we show the nonlocal spin valve measurements performed at \Vbg{}$=-60,\,-50,\,-30,\,40,$ and $50$~V, respectively, where one can see that $R_\mathrm{nl}^\mathrm{AP}$ is higher than $R_\mathrm{nl}^\mathrm{P}$ for all the backgate voltages, as in the main manuscript. The only exception is at $\Vbg{}=\,-30$~V, where \DRnl{} is smaller than the noise level ($\sim0.5$~m$\Omega$).
We also performed nonlocal spin precession measurements applying $B_x$ for the five \Vbg{} values mentioned, as shown in Figs.~\ref{FigureDifHanle}b, d, f, h, and j. The insets show the low $B_x$ region, which confirm that $R_\mathrm{nl}^\mathrm{P}$ and $R_\mathrm{nl}^\mathrm{AP}$ do not cross, with the exception of \Vbg{}$=\,-30$~V, where the in-plane signal is smaller than the noise level and both curves touch. This confirms the robustness of our measurements with \Vbg{}.

\section{Spin drift experiments across the WSe$_2$-covered region at 50 K}

\begin{figure}[tb!]
	\centering
		\includegraphics[width=\textwidth]{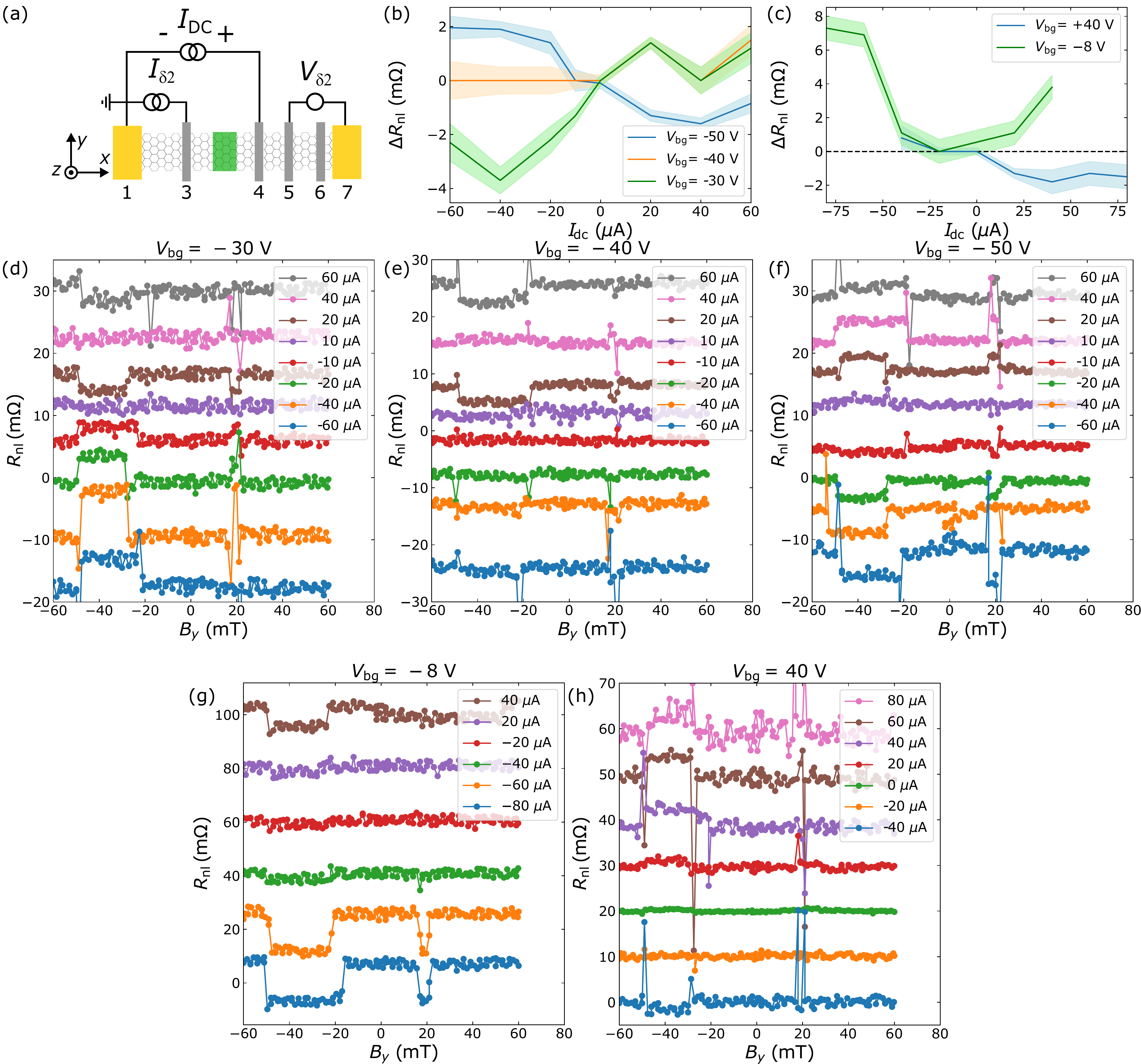}
	\caption{{Nonlocal spin valve measurements as a function of \Idc{} and \Vbg{} at 50~K.} (a) Sketch of the measurement configuration. (b) \DRnl{} vs \Idc{} at \Vbg{}$=-50$, $-40$ and $-30$~V taken from data in panels b and c. (c) \DRnl{} vs \Idc{} at \Vbg{}$=-8$ and $40$~V taken from the data in panels g and h. (d), (e), (f), (g) and (h) Nonlocal spin valve measurements as a function of \Idc{} for \Vbg{}$=-50$, $-40$, $-30$, $-8$, and $40\,$V, respectively. }
	\label{FigureSVDrift}
\end{figure}

We perform spin drift experiments across the WSe$_2$-covered BLG region using the circuit shown in Fig.~\ref{FigureSVDrift}a, with the delta current ($I_{\delta2}$) applied between contacts 3 and 1, the delta voltage ($V_{\delta2}$) measured between contacts 5 and 7, and the DC current (\Idc{}) applied between contacts 4 and 1 to induce drift.

To determine the evolution of \DRnl{} with \Idc{} and \Vbg{}, we performed nonlocal spin valve measurements in the configuration of Fig.~\ref{FigureSVDrift}a. The results from these measurements are shown in Fig.~\ref{FigureSVDrift}. Figure~\ref{FigureSVDrift}b shows the results presented in Fig.~3c of the main manuscript and the data for \Vbg{}$=-40$~V. The corresponding spin valve measurements are shown in Figs.~\ref{FigureSVDrift}d, \ref{FigureSVDrift}e, and \ref{FigureSVDrift}f and one can clearly see the crossing between $R_\mathrm{nl}^\mathrm{P}$ and $R_\mathrm{nl}^\mathrm{AP}$ at \Idc{}$= 0$ in both cases. The results obtained at \Vbg{}$=-8$~V are shown in Figs.~\ref{FigureSVDrift}c and \ref{FigureSVDrift}g and, in contrast with the other measured \Vbg{} values, \DRnl{} remains positive in the whole \Idc{} range. The increase in \DRnl{} for positive and negative \Idc{} is attributed to the opposite doping of the WSe$_2$-covered and pristine BLG regions which have comparable lengths. Figures~\ref{FigureSVDrift}f and \ref{FigureSVDrift}g show the data at \Vbg{}$=40$~V, which yields a similar trend as for \Vbg{}$=-50$~V. Negative \DRnl{} has also been observed for \Idc{}$=60\,\mu$A at \Vbg{}$=5$, $10$, and $20$~V (not shown).
As can be seen from the data, in some cases, there are large spikes in \Rnl{} when the magnetization switches. Since we do not know the origin of these one-point artifacts, we filter them out to quantify \DRnl{}.

\begin{figure}[tb!]
	\centering
		\includegraphics[width=\textwidth]{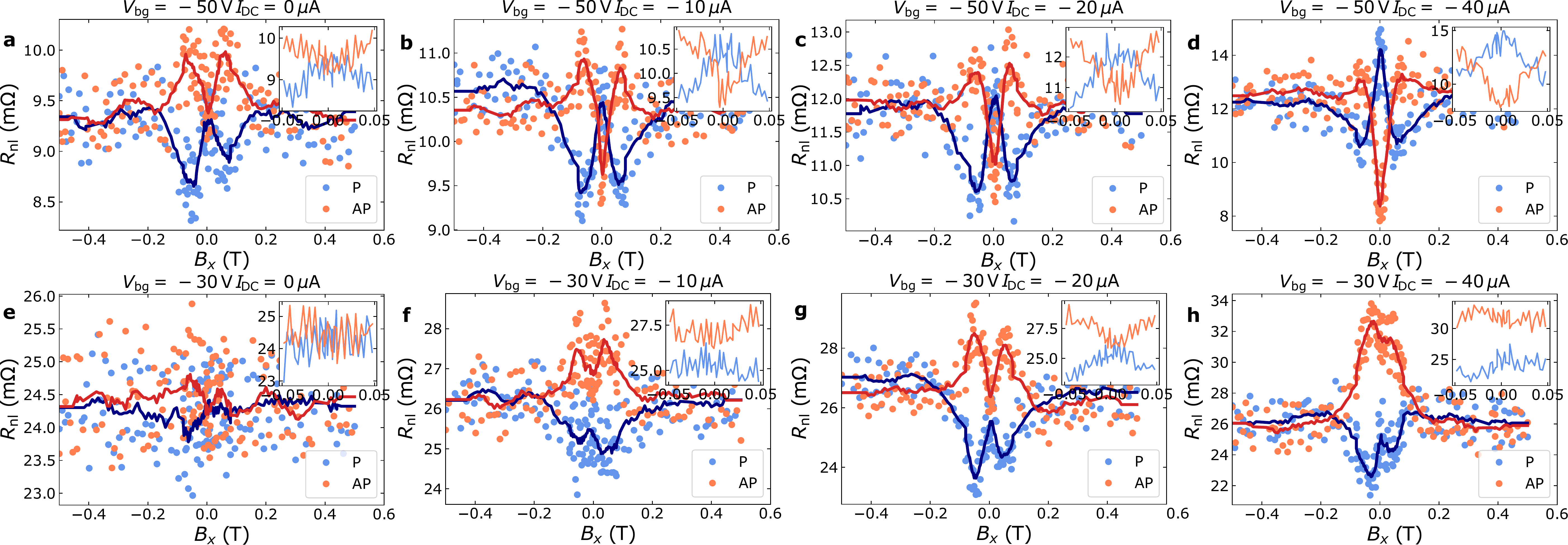}
	\caption{{Nonlocal spin precession measurements as a function of \Vbg{} and \Idc{} at 50~K.} (a-d) Spin precession measurements at \Vbg{}=~$-50$ and different \Idc{} values ($0$ to $-40\,\mu$A). (e-h) Spin precession measurements at \Vbg{}=~$-30$ and different \Idc{} values ($0$ to $-40\,\mu$A). The insets correspond to the low $B_x$ range and the solid lines have been obtained averaging a window of eleven points. }
	\label{FigureHanleDrift}
\end{figure}

The evolution of the nonlocal spin precession data with \Idc{} at \Vbg{}$=-50$ and $-30$~V is shown in Fig~\ref{FigureHanleDrift}. The top panels (Figs.~\ref{FigureHanleDrift}a-d) correspond to \Vbg{}$=-50$~V and the bottom ones (Figs.~\ref{FigureHanleDrift}e-h) correspond to the \Vbg{}$=-30$~V case. In the latter, because $R_\mathrm{nl}^\mathrm{P}<R_\mathrm{nl}^\mathrm{AP}$ at $B_x=0$, the in-plane spin signal [$\DRnl{}(B_x=0)=(R_\mathrm{nl}^\mathrm{P}(B_x=0)-R_\mathrm{nl}^\mathrm{AP}(B_x=0))/2$] decreases with \Idc{} while their absolute value increases. We also observe that the out-of-plane signal [\DRnl{}$(B_x\sim\pm100)$~mT] increases with the applied \Idc{}, as expected for standard spin drift experiments. In contrast, for \Vbg{}$=-50$~V, \DRnl($B_x=0$) increases as \Idc{} decreases while the out-of-plane spin signal changes weakly from 1.5 to 2~m$\Omega$. As a result, for \Idc{}$=-40\,\mu$A (Fig.~\ref{FigureHanleDrift}d), the in-plane spin signal is significantly larger than the out-of-plane one, as in conventional spin precession measurements in isotropic systems.

\section{Spin drift experiments across the WSe$_2$-covered region at 300~K}
\begin{figure}[tbh]
	\centering
		\includegraphics[width=\textwidth]{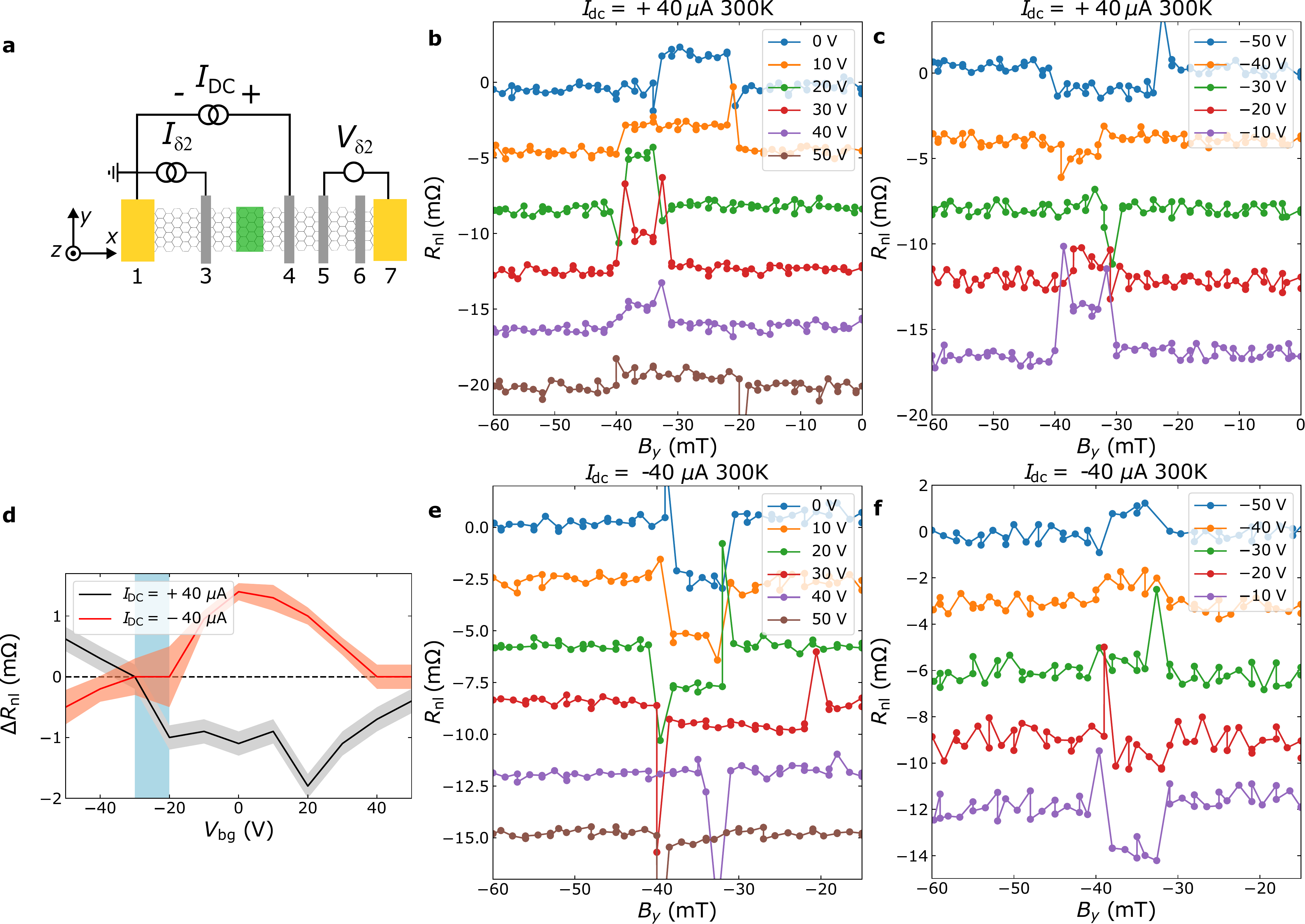}
	\caption{{Room temperature nonlocal spin valve measurements as a function of \Vbg{} for \Idc{}$=\pm40\,\mu$A.} (a) Sketch of the measurement configuration. (b), (c) Spin valve measurements obtained at \Idc{}$=+40\,\mu$A for different \Vbg{} values. The curves are offset for clarity. (e), (f) Nonlocal spin valve measurements obtained at \Idc{}$=-40\,\mu$A for different \Vbg{} values. The curves are offset for clarity. (d) \DRnl{} vs \Vbg{} for \Idc{}$=\pm40\,\mu\mathrm{A}$, as shown in Fig.~4 of the main manuscript, taken from the data in panels b, c, e, and f.}
	\label{FigureSV300K}
\end{figure}

\begin{figure}[tb]
	\centering
		\includegraphics[width=0.7\textwidth]{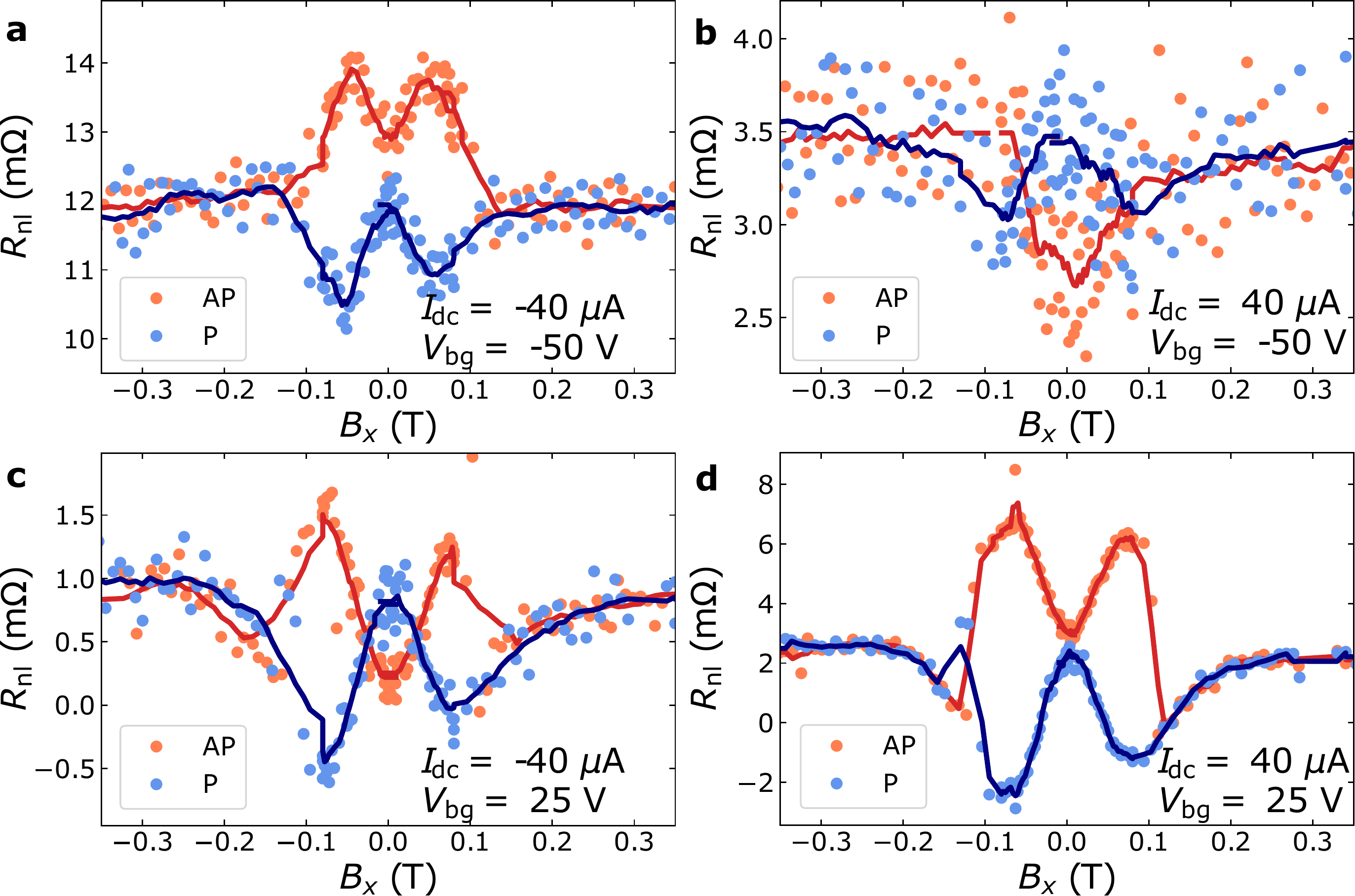}
	\caption{{Room temperature nonlocal spin precession measurements at different \Vbg{} and \Idc{} values.} \Rnl{} vs $B_x$ for \Vbg{}$=-50$~V and \Idc{}$=-40\,\mu$A (a) and \Idc{}$=+40\,\mu$A (b); \Vbg{}$=+25$~V and \Idc{}$=-40\,\mu$A (c) and \Idc{}$=+40\,\mu$A (d).}
	\label{FigureHanleDrift300K}
\end{figure}

In this section, to complement the room temperature results presented in the main manuscript, we show the results from the spin drift experiments performed at 300~K. In Fig.~\ref{FigureSV300K}, we plot the nonlocal spin valve measurements used to obtain \DRnl{} in Fig.~4 of the main manuscript. In this case, we observe that, at fixed \Idc{}, \DRnl{} changes sign once, at \Vbg{}$\approx-30$~V and decreases at \Vbg{}$=50$~V, indicating that there might be another sign change for \Vbg{}$>50$~V. Furthermore, \DRnl{} is reversed by changing the sign of \Idc{}, as at 50~K.
We also measured the spin precession data at \Vbg{}$=-50$ and $+25$~V and \Idc{}$=\pm40\,\mu$A. The results are shown in Fig.~\ref{FigureHanleDrift300K} and show that, while the in-plane spin signal reverses sign, the out-of-plane spin signal remains positive in all the cases. Note that, despite the significantly smaller signal in Fig.~\ref{FigureHanleDrift300K}b, $R_\mathrm{nl}^\mathrm{P}$ clearly crosses $R_\mathrm{nl}^\mathrm{AP}$ at $B_x\approx\pm80$~mT.
 
\FloatBarrier

\section{Spin drift experiments at the pristine graphene region at 300~K}\label{SectionDriftOuter}
\begin{figure}[tbh!]
	\centering
		\includegraphics[width=\textwidth]{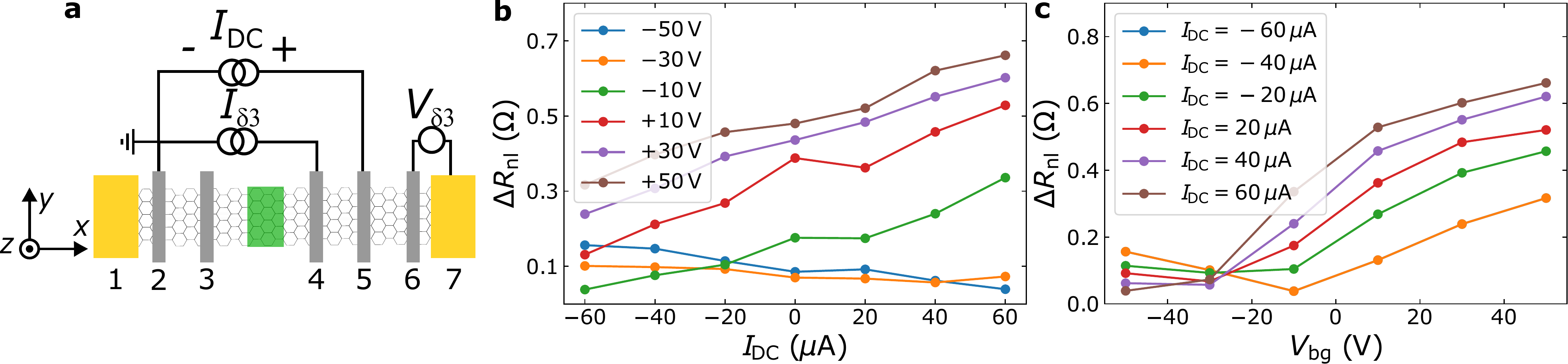}
	\caption{{Room temperature spin drift experiments in the pristine BLG region.} (a) Sketch of the measurement configuration. (b) \Idc{}-dependence of \DRnl{} at different \Vbg{} values. (c) \Vbg{}-dependence of \DRnl{} at different \Idc{} values. }
	\label{FigureDriftOuter300K}
\end{figure}

To confirm that the anomalous \Idc{}-dependence reported here is caused by the WSe$_2$-covered region, we performed spin drift experiments in the pristine BLG region.
The measurement configuration is shown in Fig.~\ref{FigureDriftOuter300K}a. The spins are injected by applying a delta current $I_{\delta3}$ between contacts 4 and 2. Note that, even though contact 2 may inject some spins, most of them relax in the WSe$_2$-covered BLG region and no additional switches were observed in the nonlocal spin valve measurements. The nonlocal signal ($V_{\delta3}$) is measured between contacts 6 and 7, that is a non-magnetic electrode. Finally, \Idc{} is applied between contacts 5 and 2 to induce drift. In Fig.~\ref{FigureDriftOuter300K}b, we show the evolution of \DRnl{} with \Idc{} for different values of \Vbg{}. We observe that, for \Vbg{}=$-50$ and $-30$~V, \DRnl{} decreases as \Idc{} increases. Looking at Fig.~\ref{FigureRsq}c, one realizes that the CNP of the pristine BLG region at 300~K occurs at \Vbg{}$\approx-27$~V, thus, the channel is hole doped for \Vbg{}$=-50$ and $-30$~V. In this case, the application of a positive \Idc{} makes holes drift from contact 5 to contact 2, opposing spin transport. Hence, \DRnl{} is expected to decrease as \Idc{} increases. In contrast, for \Vbg{}=$-10,\,10,\,30,$ and $50$~V, \DRnl{} increases with \Idc{}. This observation is consistent with the fact that electrons in the channel propagate in the opposite direction of \Idc{}, flowing from contact 2 to 5 for \Idc{}$>0$, and giving rise to an enhancement of \DRnl{} with \Idc{} \cite{jozsa2008,ingla2016}. These results imply that the anomalous \Idc{}-dependence of \DRnl{} measured across the WSe$_2$-covered BLG channel is induced by the unconventional spin transport in the latter.

\section{Reproducibility}

\begin{figure}[tbh!]
	\centering
		\includegraphics[width=0.7\textwidth]{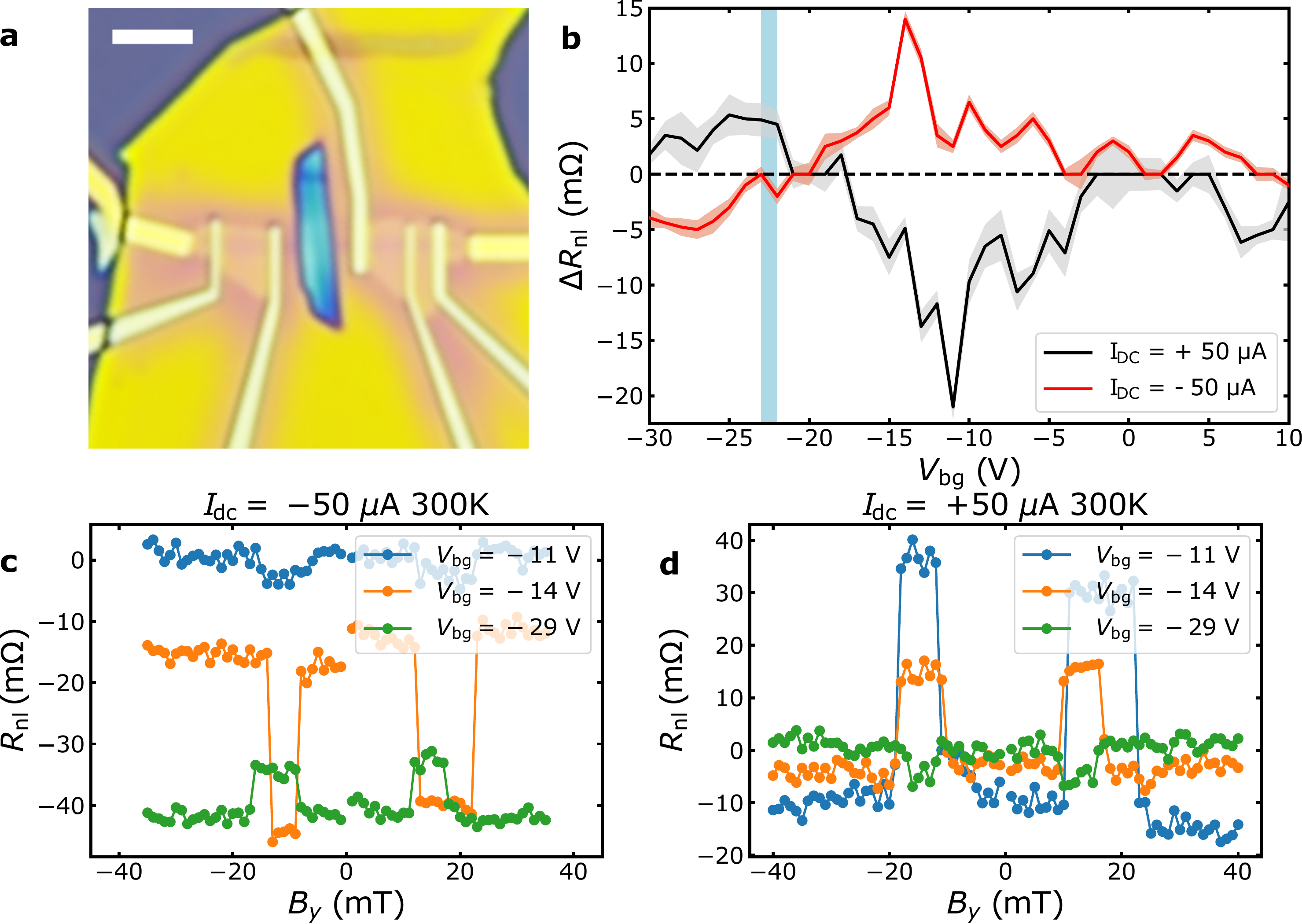}
	\caption{{Spin drift experiments in device 2 at 100 K.} (a) Optical image of device 2. The vertical WSe$_2$ flake (blue) and Co electrodes (light grey) are visible below the hBN flake (yellow) used to cap the device. The horizontal graphene flake is barely visible as a darker strip underneath the electrodes and the TMD flake. The scale bar is 3~$\mu$m. For the spin drift measurements, the same configuration as in Fig.~\ref{FigureSV300K}a was used. (b) \Vbg{}-dependence of \DRnl{} for \Idc{}$ = \pm50\,\mu$A and $I_{\delta2} = 50\,\mu$A. The blue vertical strip represents the CNP of the WSe$_2$-covered region. The light grey and red areas represent the experimental error range calculated using the standard deviation associated with the statistical average of $R_\mathrm{nl}^\mathrm{P}$ and $R_\mathrm{nl}^\mathrm{AP}$. (c) and (d) Nonlocal spin valve measurements at \Idc{}$=-/+50\,\mu$A, respectively and at \Vbg{}$=-11$, $-14$, and $-29$~V.}
	\label{FigureSample2}
\end{figure}

In addition to the device reported until this point (device 1), we prepared another sample (device 2) using the same recipe (reported in the Methods section of the main manuscript) and obtained very similar results, showing the robustness of our findings in BLG/WSe$_2$ van der Waals heterostructures. The optical microscope image of the device and the \Vbg{}-dependence of \DRnl{} for two different \Idc{} values are shown in Fig.~\ref{FigureSample2}.

As in Fig.~2d and 3b of the main manuscript, the sign of \DRnl{} in the non-local spin valve configuration reverses close to the CNP of the BLG region proximitized with WSe$_2$. Sweeping \Vbg{} controls the amplitude and the sign of the spin signal, as shown in Fig.~\ref{FigureSample2}b. Due to the width of the WSe$_2$-covered graphene region being 1.4 $\mu$m (narrower than in device 1), the amplitude of the spin signal is roughly ten times larger than in device 1, and the signal-to-noise ratio of \DRnl{} is higher.
As shown in the inset of Fig.~1b of the main manuscript, when the width of the WSe$_2$ flake is designed so that the position where $\mu_{sy}$ changes sign ($x_0$) lies within the proximitized region, the sign of \DRnl{} can be controlled by the applied \Idc{}. Indeed, for the full range of \Vbg{} values, reversing the sign of \Idc{} reverses the sign of the spin signal. This shows that we have achieved the electrical control of spin reversal in two independent ways without the need for a magnetic field in two different BLG/TMD heterostructures.
\FloatBarrier
\section{Influence of the annealing temperature  and WSe$_2$ width on the proximity-induced SOC}
\begin{figure}[tbh!]
	\centering
		\includegraphics[width=\textwidth]{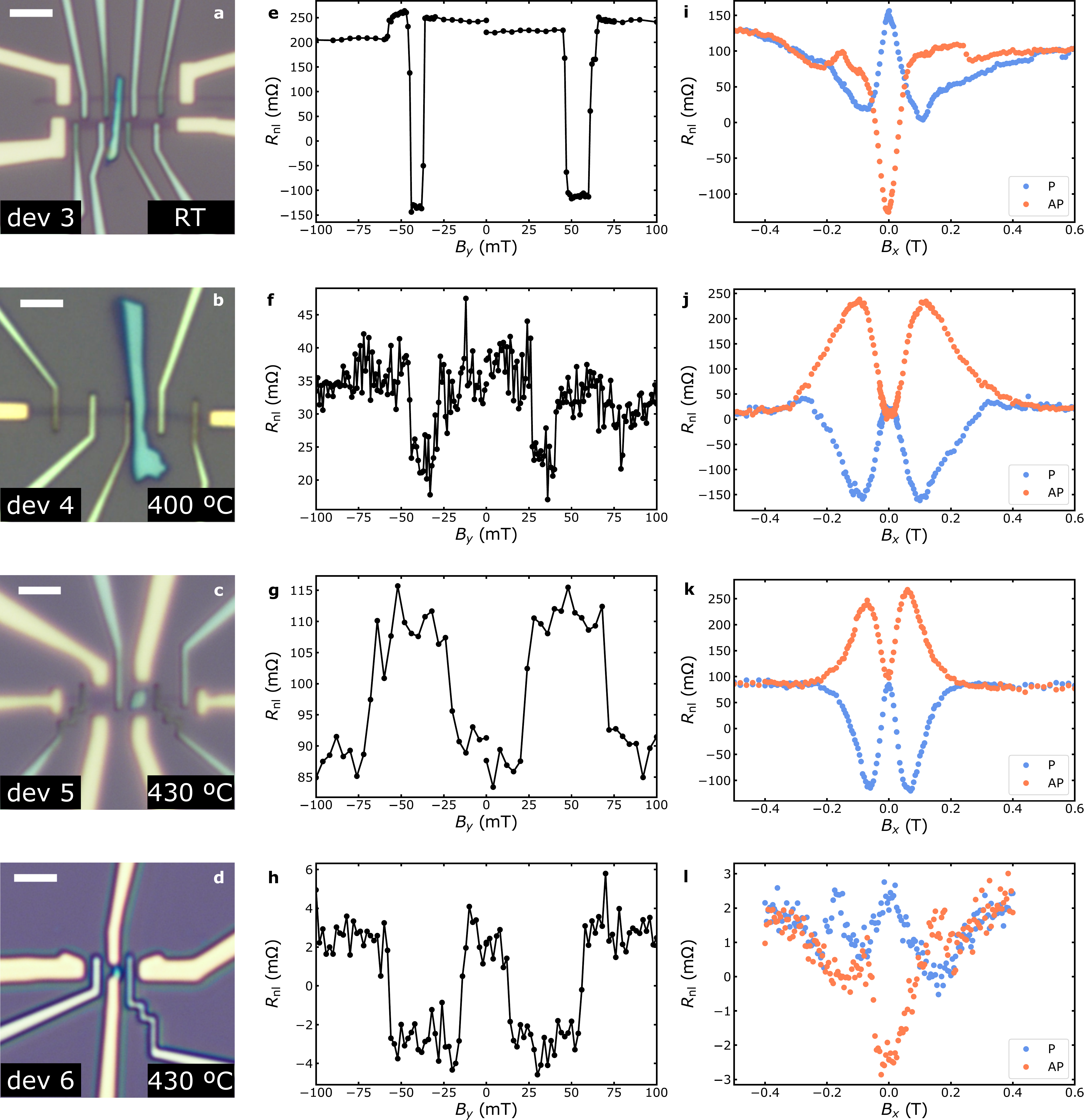}
	\caption{{The influence of SOC strength on the spin transport.} (a)-(d) Optical image of additional measured devices. The scale bar is 3~$\mu$m. (e)-(h) Nonlocal spin valve measurement across the WSe$_2$-covered graphene region as a function of the magnetic field applied along $y$ ($By$). (i)-(l) Nonlocal spin precession measurements with the magnetic field applied along $x$ ($B_x$) for the parallel (P) and antiparallel (AP) configurations. The measurement temperatures are 10~K for (e) and (i); 100~K for (f), (g), (j) and (k) and 300~K for (h) and (l).}
	\label{FigureAnnealingTemp}
\end{figure} 
The data of four additional devices are shown here to illustrate the influence of the annealing temperature, and thus SOC strength, on the spin transport. Devices 3 to 6, shown in Figs.~\ref{FigureAnnealingTemp}a-d, are fabricated following the recipe of device 1 and 2 (see the Methods section of the main manuscript) with small variations. None of the samples were capped with Au and hBN, which does not influence the measurements but makes the Co electrodes more prone to oxidation during transfer to the cryostat.
The strength of the proximity-induced SOC directly depends on the interface between graphene and the TMD flake. Some van der Waals heterostructures have a self-cleaning mechanism. Hence, after annealing, the contamination between the layers clusters in pockets and the remaining contact areas are cleaned \cite{purdie2018}. Accordingly, device 3, which was not annealed, shows a conventional positive nonlocal spin valve signal with a large amplitude (Fig.~\ref{FigureAnnealingTemp}e) as for pristine graphene samples. Also, the spin precession data (Fig.~\ref{FigureAnnealingTemp}i) has a similar shape as for isotropic systems, since a small proximity-induced SOC also leads to a negligible anisotropy in the spin transport. Along these lines, device 4, which was annealed for 1 h as all other devices, but at a lower temperature (400 $^\circ$C), shows anisotropic spin precession (Fig.~\ref{FigureAnnealingTemp}j) but no negative sign of \DRnl{} (Fig.~\ref{FigureAnnealingTemp}f). Here, the annealing led to a significant proximity-induced SOC but, due to the lower annealing temperature, it did not reach the strong SOC regime. 
We suspect that the dominating factor for the resulting strength of the induced SOC after the annealing is the temperature of the process and that the duration and the pressure only play a secondary role. This seems to agree with findings in the fabrication of other graphene/TMD heterostructures \cite{benitez2018}. 
The fabrication of devices 5 and 6 included an additional electron beam lithography and reactive ion etching step to structure the graphene flake into a double H-bar shape before fabricating the Au contacts and Co electrodes. This enables us to also measure the spin-to-charge conversion by spin Hall effect in graphene due to the proximity-induced SOC as reported in Ref.~\onlinecite{herling2020}. Both samples were annealed at 430~$^\circ$C for 1~h but have different widths of the WSe$_2$ flake, 790 and 250~nm, respectively, as obtained from scanning electron microscopy images. The wider flake leads to a longer transport in proximitized graphene, reaching the region were \DRnl{} has a negative sign (in Fig.~1b of the main manuscript, we estimate this region for a width between 750 and 1750 nm for \tiv{}=156 fs). Accordingly, the diffusive spin transport data from device 5 (Figs.~\ref{FigureAnnealingTemp}g and k) is analogous to the one shown in Figs.~2b and d. For the narrower device 6, the measured data resembles the typical behavior for spin precession in isotropic media. This is caused by the fact that the WSe$_2$ width is smaller than the in-plane spin lifetime \cite{leutenantsmeyer2018}, reducing the total in-plane spin relaxation.

\section{Modeling details}
\begin{table}[t]
    \caption{Modelling parameters.}
        \begin{ruledtabular}
        \begin{tabular}{c c c c c c}
            \renewcommand{\arraystretch}{2}
            $\lambda_\mathrm{VZ}$ & $\Omega_\mathrm{VZ}$ & $D_s$&$\tau_s^{\perp}$& $W_\mathrm{TMD}$&$W_\mathrm{gr}$\\
			(meV)&(s$^{-1}$)	&(m$^2$/s)&(ps)	&($\mu$m)&($\mu$m)\\
            \hline
             0.595 & 1.808$\times10^{12}$&0.02 & 30 & 5 & 0.5\\
    \end{tabular}
    \end{ruledtabular}
    \label{TableS2}
\end{table}

The plot shown in Fig.~1b, that confirms that our explanation is consistent with the current understanding of BLG/TMD heterostructures, is realized using spin transport calculations. In particular, assuming $\tp{}\Omega_\mathrm{R}\ll1$ (weak SOC regime for \tper{}), we use the drift-diffusion equations for each valley separately \cite{yue2016}:
\begin{equation}
\frac{d\vec{\mu}_s^\mathrm{K}}{dt}=D_s\frac{d^2\vec{\mu}_s^\mathrm{K}}{dx^2}-v_d\frac{d\vec{\mu}_s^\mathrm{K}}{dx}-\frac{\vec{\mu}_s^\mathrm{K}}{\overline{\tper{}}}+\vec{\Omega}_\mathrm{VZ}\times\vec{\mu}_s^\mathrm{K}-\frac{\vec{\mu}_s^\mathrm{K}-\vec{\mu}_s^\mathrm{K'}}{2\tiv{}}
\label{EquationBloch1}
\end{equation}
\begin{equation}
\frac{d\vec{\mu}_s^\mathrm{K'}}{dt}=D_s\frac{d^2\vec{\mu}_s^\mathrm{K'}}{dx^2}-v_d\frac{d\vec{\mu}_s^\mathrm{K'}}{dx}-\frac{\vec{\mu}_s^\mathrm{K'}}{\overline{\tper{}}}-\vec{\Omega}_\mathrm{VZ}\times\vec{\mu}_s^\mathrm{K'}-\frac{\vec{\mu}_s^\mathrm{K'}-\vec{\mu}_s^\mathrm{K}}{2\tiv{}}
\label{EquationBloch2}
\end{equation}
where, $\vec{\mu}_s^\mathrm{K(K')}=(\mu_{sx}^\mathrm{K(K')},\mu_{sy}^\mathrm{K(K')},\mu_{sz}^\mathrm{K(K')})$ is the $x$, $y$, $z$-polarized spin accumulation in valley K(K'). The first term at the right hand side accounts for spin diffusion with a diffusivity $D_s$ and the second term introduces spin relaxation with a rate  
\begin{equation}
(\overline{\tper{}})^{-1}=
\begin{pmatrix}
(2\tper{})^{-1} & 0 & 0\\
0 & (2\tper{})^{-1} & 0\\
0 & 0 & (\tper{})^{-1}\\
\end{pmatrix},
\label{EquationBloch3}
\end{equation}
 that induces an anisotropy of 1/2.
The third term induces spin precession around the VZ-SOF $\vec{\Omega}_\mathrm{VZ}=(0,0,\Omega_\mathrm{VZ})$ and the last one accounts for intervalley scattering at a rate $\tiv{}^{-1}$.
{Note that VZ-SOF fluctuations may lead to extra in-plane spin relaxation \cite{zutic2004} which is not included in the model but should not lead to net spin precession.} 
\subsection{Weak SOC regime}
To confirm that the output of our model agrees with the existing theory in the weak SOC regime (\tiv{}$\ll\tau_\mathrm{VZ}$), we have solved Eqs.~\ref{EquationBloch1} and \ref{EquationBloch2} in the steady state ($d\vec{\mu}_s^\mathrm{K(K')}/dt=0$) using the parameters displayed in Table~\ref{TableS2}.
This operation is performed numerically as a boundary value problem using Python's Scipy package. We solve the problem in the range $0<x<5\,\mu$m and, to set the boundary conditions, we define the spin current as follows:
\begin{equation}
\vec{I}_s^\mathrm{K(K')}=\frac{W_\mathrm{gr}}{e\Rsq{}}\left(-\frac{d\vec{\mu}_s^\mathrm{K(K')}}{dx}+\frac{v_d}{D_s}\vec{\mu}_s^\mathrm{K(K')}\right),
\end{equation}  
where we use \Rsq{}$=1$~k$\Omega$, and $W_\mathrm{gr}=0.5\,\mu$m. At $x=0$, we assume that $I_{sy}^\mathrm{K}(x=0)=I_{sy}^\mathrm{K'}(x=0)=P_i/4$, with $P_i=0.1$. At $x=W_\mathrm{TMD}=5\,\mu$m, we assume that $I_{sy}^\mathrm{K}(x=W_\mathrm{TMD})=I_{sy}^\mathrm{K'}(x=W_\mathrm{TMD})=0$. 
In the weak SOC regime, the spin accumulation ($\mu_{sy}=\mu_{sy}^\mathrm{K}+\mu_{sy}^\mathrm{K'}$) decays exponentially and can be written using 
\begin{figure}[tbh!]
	\centering
		\includegraphics[width=0.7\textwidth]{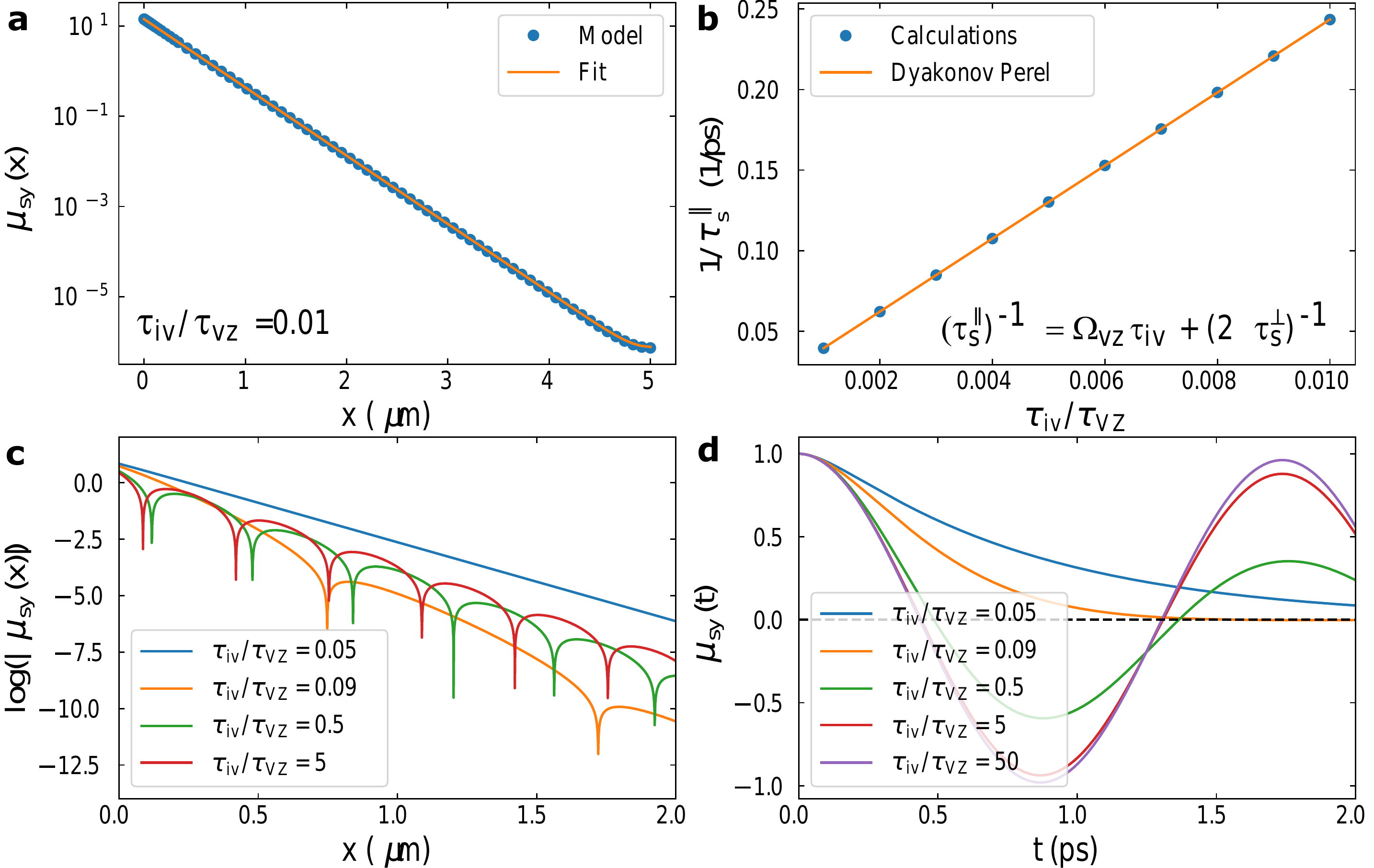}
	\caption{{Output results of the two-valley model at \Idc{}$=0$.} (a) $x$-dependence of $\mu_{sy}$ obtained from our model and its fit to Eq.~\ref{EquationMuS}. (b) In-plane spin relaxation rate as a function of \tiv{} in the weak SOC regime. (c) In-plane spin accumulation as a function of $x$ for different values of \tiv{}. (d) In-plane spin accumulation as a function of time after an initial pulse at time $t=0$ for different values of \tiv{}.}
	\label{FigureModel}
\end{figure} 
 
\begin{equation}
\mu_{sy}=A\exp(x/\lambda_s)+B\exp(-x/\lambda_s),
\label{EquationMuS}
\end{equation}
where $\lambda_s=\sqrt{D_s\tau_s}$.
\begin{equation}
A=\frac{eP_i\Rsq{}\lambda_s}{2W_\mathrm{gr}}\frac{1}{1-\exp(2W_\mathrm{TMD}/\lambda_s)},
\end{equation}
and
\begin{equation}
B=A\exp(2W_\mathrm{TMD}/\lambda_s)
\end{equation}
 are obtained from the boundary conditions for $I_s^\mathrm{K(K')}$.
To determine the effective spin lifetime, we fit the results from the two-valley model to Eq.~\ref{EquationMuS} with $\lambda_s$ as the only fitting parameter (see Fig.~\ref{FigureModel}a). $\tau_s$ is obtained using $\tau_s=\lambda_s^2/D_s$. Next, we plot the in-plane spin relaxation rate ($1/\tau_s^\parallel$) as a function of \tiv{}, which we have normalized to $\tau_\mathrm{VZ}$. Figure~\ref{FigureModel}b shows that $1/\tau_s^\parallel$ is linear with respect to \tiv{}, as expected in the weak SOC regime. The orange line in Fig.~\ref{FigureModel}b is a linear fit $\tpar{}^{-1}=C\tiv{}+D$, where $C=3.262\times10^{24}\pm6\times10^{20}$~s$^{-2}$ and $D=1.671\times10^{10}\pm1\times10^{7}$~s$^{-1}$ are the fitting coefficents. The inset text is the spin relaxation rate in the weak SOC (Dyakonov-Perel) regime \cite{cummings2017}. We confirm the consistency of our model using that $C=\Omega_\mathrm{VZ}^2$ and $D=(2\tper{})^{-1}$ and obtain \tper{}$=29.9\pm0.04$~ps and $\Omega_\mathrm{VZ}=(1.806\times10^{12}\pm2\times10^{8})$~s$^{-1}$. These results agree well with the input parameters shown in Table~\ref{TableS2}.  

\subsection{Strong SOC regime}
\begin{figure}[tb!]
	\centering
		\includegraphics[width=\textwidth]{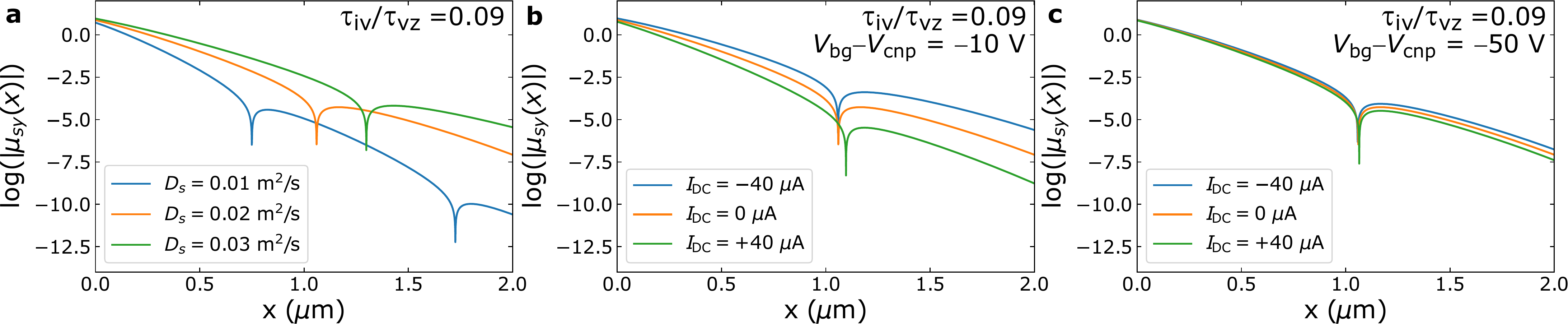}
	\caption{{Effect of changing $D_s$ and $n$ on $\mu_{sy}$. (a) $\mu_{sy}$ vs $x$ at \Idc{}$=0$ and $D_s=0.01,\,0.02$, and $0.03$~m$^2$/s. $\mu_{sy}$ vs $x$ at \Idc{}$=-40\,,0,$ and $40\,\mu$A for \tiv{}=0.09\tvz{}, \Ds{}$=0.02$~m$^2$/s and (b) for \Vbg{}-\Vcnp{}$=-10$~V and (c) for \Vbg{}-\Vcnp{}$=-50$~V.}}
	\label{FigureModel2}
\end{figure}
Once we have confirmed that our model works as expected in the weak SOC regime, we look at the strong SOC regime. For this purpose, we calculated $\mu_{sy}(x)$ for $0.05<\tiv{}/\tvz{}<5$ (see Fig.~\ref{FigureModel}c). We observe that, for $\tiv{}/\tvz{}=0.05$, $\mu_{sy}(x)$ does not change sign for $0<x<2\,\mu$m. In contrast, for $0.09<\tiv{}/\tvz{}<5$, $\mu_{sy}(x)$ oscillates with $x$ at a frequency that depends on \tiv{}. To understand if the change in the oscillation length comes from a change in the precession frequency around the SOFs or just a change in the spin lifetime, we have calculated $\mu_{sy}(t)$ by assuming homogeneous spin injection ($\frac{d^2\vec{\mu}_s^\mathrm{K(K')}}{dx^2}=\frac{d\vec{\mu}_s^\mathrm{K(K')}}{dx}=0$) at all times, and $\mu_{sy}(t=0)=1$. Then we calculate $\mu_{sy}(t)$ using Eqs.~\ref{EquationBloch1} and \ref{EquationBloch2} and integrating them as ordinary differential equations with the input parameters shown in Table~\ref{TableS2}.
 The results from this operation for $0.05<\tiv{}/\tvz{}<50$ are shown in Fig.~\ref{FigureModel}d. We observe that, for \tiv{}/\tvz{}$=0.05$ and $0.09$, there are no clear oscillations. For \tiv{}/\tvz{}$=0.09$ the reason for this low signal is the fast decay of $\mu_{sy}(t)$, that we attribute to strong Dyakonov-Perel dephasing. In contrast, for $0.5<\tiv{}/\tvz{}<50$, we see clear oscillations at a frequency that changes with \tiv{} at $0.5<\tiv{}/\tvz{}<5$. As expected from the fact that the spins can complete several rotations between intervalley scattering events, for \tiv{}/\tvz{}=50 the frequency is the same as for \tiv{}/\tvz{}=5 and the only difference between these curves is that the former has a slightly slower decay.
 
 To disentangle the origin of the sign change of the in-plane spin signal with \Vbg{} reported in the main manuscript, one has to consider the different parameters which are changing with \Vbg{}. These parameters are $\tiv{}$, $D_s$,  $n$ and \tper{}. The role of \tiv{} is shown in Fig.~\ref{FigureModel}c. $D_s$ changes the spin diffusion time (Fig.~\ref{FigureModel2}a), $n$ modulates the effect of drift via $v_d=\Idc{}/(W_{gr}ne)$, and \tper{} gives rise to additional spin relaxation. We found that \tper{} does not modify $\mu_{sy}(x)$ significantly unless the change in \tper{} is dramatic (from 5 to 60~ps), at odds with experimental results \cite{benitez2018}. In contrast, as shown in Fig.~\ref{FigureModel2}a, $D_s$ has a very strong influence on $\mu_{sy}$ and is the most likely cause of the observed sign change with \Vbg{}.
 Changing \Vbg{} ($n$) while keeping the other parameters fixed leads to the modification of the effect of drift, as shown in Fig.~\ref{FigureModel2}b and \ref{FigureModel2}c and expected from the fact that $v_d=\Idc{}/(W_{gr}en)$. Accordingly, $d\DRnl{}/d\Idc{}$ is expected to change magnitude with $n$ but not sign, as observed in the data.

The conclusion from this section is that the sign reversal of \DRnl{} with \Vbg{} and \Idc{}$\neq0$ is most likely caused by the change in $D_s$, which we obtain here using $D_c$. However, we cannot discard a change of \tiv{} with \Vbg{}, which can also lead to the sign reversal of \DRnl{}.
\subsection{\Idc{}-dependence of \DRnl{}.}
\begin{figure}[tb!]
	\centering
		\includegraphics[width=\textwidth]{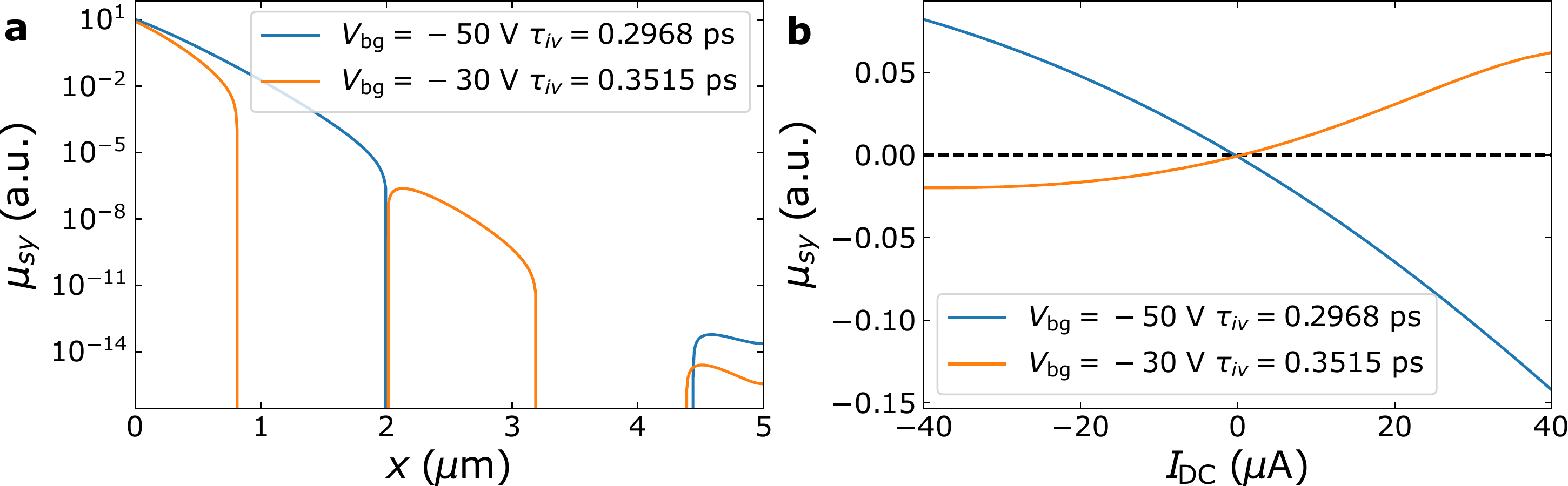}
	\caption{$\mu_{sy}$ sign change with \Idc{} at different \Vbg{}. (a) $\mu_{sy}$ as a function of $x$ at \Idc{}=0 and (b) as a function of \Idc{} for $\Vbg{}=-30$ and $-50$~V.}
	\label{FigureIdcDepDRnl}
\end{figure}

Here we discuss the measured \Idc{}-dependence of \DRnl{} in terms of our two-valley model.
The most relevant feature of the data displayed in Figs.~\ref{FigureSVDrift} to \ref{FigureHanleDrift300K} is that \DRnl{} changes sign near \Idc{}=0. This result is surprising at a first glance, since it contrasts with the spin drift experiments shown in the pristine graphene region (Section~\ref{SectionDriftOuter} and Refs.~\onlinecite{jozsa2008, ingla2016}). In this section, we show that the \DRnl{} vs \Idc{} results measured across the TMD-covered graphene region are compatible with the model described by Equations~\ref{EquationBloch1}-\ref{EquationBloch3}. For this purpose, we evaluate the spin accumulation at $x=2$~$\mu$m as a function of \Idc{}. The outcome of this calculation, using the parameters from Table~\ref{TableS2}, is shown in Fig.~\ref{FigureIdcDepDRnl}.

By tuning \tiv{}, it is possible to calibrate the position were the sign reversal of \musy{} occurs and bring it to $x=2$~$\mu$m (see Fig.~\ref{FigureIdcDepDRnl}a). 
In this case, a small positive (negative) \Idc{} can increase (decrease) the spin transport time and lead to a positive (negative) \DRnl{} as reported here for \Vbg{}$=-30$~V and viceversa for \Vbg{}$=-50$~V (see Fig.~\ref{FigureIdcDepDRnl}).
Note that, even though the sign reversal of \DRnl{} with \Idc{} is compatible with the model, it occurs only for specific values of \tiv{}, which correspond to well defined spin precession frequencies. This is not necessarily the case for all values of \Vbg{} so we would expect that, at other \Vbg{}, \DRnl{} may be sizeable for \Idc{}=0, contrary to what is observed experimentally.

The results from Fig.~\ref{FigureIdcDepDRnl} show that the magnitude of \DRnl{} can increase when applying a positive and a negative \Idc{}, a feature which is not possible in the weak SOC regime. Note that the simulation does not include the effect of drift in the pristine graphene region that is expected to enhance the signal for negative \Idc{} and decrease the signal for positive \Idc{}.

We can list different features of the \Idc{} and \Vbg{}-dependence of \DRnl{} that strongly indicate that the measured dependence is inherent to the spin transport across the WSe$_2$-covered region and not caused by a spurious effect from the spin drift measurement setup:
\begin{itemize}
\item The \Idc{}-dependence of \DRnl{} at the pristine graphene region shows the expected trend for graphene (Section~\ref{SectionDriftOuter} and Refs.~\onlinecite{jozsa2008, ingla2016}). This means that the electronic setup, that is described in Section~\ref{SectionElMeas}, is well calibrated to separate \Idc{} from \Idelta{}.
\item The sign change of \DRnl{} with \Idc{} does not occur at all \Vbg{} values. For \Vbg{} = $-8$~V (Figs.~\ref{FigureSVDrift}c and \ref{FigureSVDrift}g), \DRnl{} is positive for both positive and negative \Idc{} and, for \Vbg{} $=\,-40$~V (Figs.~\ref{FigureSVDrift}b and \ref{FigureSVDrift}e), a measurable spin signal is only observed for positive \Idc{}. Such dependence cannot occur in the weak SOC regime but is expected from the model described by Equations~\ref{EquationBloch1}-\ref{EquationBloch3}.
\item  \DRnl{} changes sign multiple times when sweeping \Vbg{} at fixed \Idc{} (Fig.~3d of the main manuscript). Hence, it is unlikely that the sign change is due to a simple modification of the charge current path with \Vbg{}, as \Rsq{} in both the pristine and WSe$_2$-covered graphene regions have very similar \Vbg{}-dependences (Fig.~\ref{FigureRsq}).
\item The spin precession data shown in Figs.~\ref{FigureHanleDrift} and \ref{FigureHanleDrift300K} shows that \DRnl{} at $B_x\approx\pm0.1$~T, which corresponds to the shoulders where the spins travel across the WSe$_2$-covered region out-of-plane, does not change sign with \Idc{} or \Vbg{}. This is consistent with the fact that out-of-plane spins are not affected by the valley-Zeeman SOFs.
\end{itemize}
Finally, we note that, unlike that of Ref.~\onlinecite{offidani2018}, our model does not represent a fully relativistic description of the problem and may not be able to fully predict the expected dependencies of \DRnl{}.
\subsection{Influence of the charged impurities on spin transport in the strong SOC regime}
In the weak SOC regime, charge impurities were shown not to affect spin relaxation in graphene while having a significant influence on the electron mobility \cite{han2012}.
In contrast, in the strong SOC regime, the spin precession angles depend on the spin diffusion/drift time, that is dictated by charge transport, and \tiv{} (see Figs.~\ref{FigureModel}c and \ref{FigureModel}d). 
Thus, the charge traps at the interface between BLG and WSe$_2$ may play a role in the backgate dependence of the spin signal. Even though this contribution is already included in the mobility and contributes to the diffusivity, it cannot be separated from other mechanisms limiting the mobility in our devices. 
\tiv{} is not affected by charge impurities because Coulomb interactions have a long range.
\section{Tight binding calculations}
The band structure shown in Figs.~2e and 2f of the main manuscript has been obtained using the Hamiltonian described in Ref.~\onlinecite{zollner2020}.
Here, for the sake of completeness, we show the Hamiltonian, that has two different components:
\begin{equation*}
H_\mathrm{orb} = 
\begin{pmatrix}
\Delta+V & \gamma_0f(k) & \gamma_4f^*(k) & \gamma_1 \\
\gamma_0f^*(k) & V & \gamma_3f(k) & \gamma_4f^*(k) \\
\gamma_4f(k)  & \gamma_3f^*(k)  & -V & \gamma_0f(k)  \\
\gamma_1 & \gamma_4f(k) & \gamma_0f^*(k) & \Delta-V 
\end{pmatrix}
\otimes s_0,
\end{equation*}
accounting for the orbital effects and
\begin{equation*}
H_\mathrm{SOC}+H_\mathrm{R} = 
\begin{pmatrix}
\tau \liA1{}s_z & i(\lo{}+2\lr{})s_{-}^\tau & 0 & 0 \\
-i(\lo{}+2\lr{})s_{+}^\tau & -\tau\liB1{}s_z & 0 & 0 \\
0 & 0 & \tau\liA2{}s_z & -i(\lo{}-2\lr{})s_{-}^\tau  \\
0 & 0 & i(\lo{}-2\lr{})s_{+}^\tau & -\tau\liB2{}s_z
\end{pmatrix}
,
\end{equation*}
accounting for the spin-orbital effects. The latter is written on the basis elements: $\ket{C_\mathrm{A1},\uparrow}$, $\ket{C_\mathrm{A1},\downarrow}$, $\ket{C_\mathrm{B1},\uparrow}$,  $\ket{C_\mathrm{B1},\downarrow}$, $\ket{C_\mathrm{A2},\uparrow}$, $\ket{C_\mathrm{A2},\downarrow}$, $\ket{C_\mathrm{B2},\uparrow}$, and $\ket{C_\mathrm{B2},\downarrow}$, where A1(2) and B1(2) correspond to the A and B sites on the bottom(top) layer, respectively. $\gamma_{0,1,3,4}$ are the tight-binding parameters\cite{zollner2020} (Table~\ref{TableS3}) and $f(k)=\exp\left(\frac{ik_ya}{\sqrt{3}}\right)+2\exp\left(-\frac{ik_ya}{2\sqrt{3}}\right)\cos\left( \frac{k_xa}{2}\right)$ the nearest-neighbour structural function where $k=(k_{x}, k_{y})$ is a reciprocal space vector and $a=0.246$~nm the lattice constant. $s_{\pm}^{\tau}=(s_x\pm i\tau s_y)/2$ where $\tau=+(-)1$ in valley K(K'). The band structure of Fig.~2e has been calculated by finding the eigenvalues and eigenvectors of the Hamiltonian $H_\mathrm{tb}=H_\mathrm{orb}+H_\mathrm{SOC}+H_\mathrm{R}$ for valley K.
\begin{table}[t]
    \caption{Tight binding parameters in eV.}
        \begin{ruledtabular}
        \begin{tabular}{c c c c c c}
            \renewcommand{\arraystretch}{2}
            $\g0{}$ & $\g1{}$&$\g3{}$&$\g4{}$& $\Delta$&$V$\\
            \hline
             2.453 & 0.372 &$-$0.27&$-$0.162&$-$10.208$\times10^{-3}$&0\\
             \hline
             \hline
             \liA1{} & \liA2{}&\liB1{}&\liB2{} & \lo{}& \lr{}\\
            \hline
             0 & 1.07$\times10^{-3}$ &0&$-$1.179$\times10^{-3}$&0&0\\
    \end{tabular}
    \end{ruledtabular}
    \label{TableS3}
\end{table}
\section{Effect of a perpendicular electric field on the spin texture of bilayer graphene}
When its inversion symmetry is broken by a perpendicular electric field, bilayer graphene opens a gap and displays an out-of-plane spin splitting near the charge neutrality point that has the same symmetry as the valley-Zeeman SOC \cite{konschuh2012}. However, the magnitude of this SOC is 12~$\mu$eV, leading to a precession period \tso{}=172~ps. %
Such long spin precession period, which increases quickly as the carrier density is tuned far from the charge neutrality point, is already significantly longer than typical intervalley scattering times in bilayer graphene on SiO$_2$, that are in the range of 10~ps \cite{gorbachev2007}.
Accordingly, recent experimental works have shown that spin transport in bilayer graphene remains in the weak SOC regime, with long spin lifetimes in the ns range, even when a perpendicular electric field is present near the charge neutrality point \cite{leutenantsmeyer2018, xu2018}.
We conclude that the effect of the perpendicular electric field alone (without the SOC induced by the WSe$_2$) cannot account for the observed spin precession in bilayer graphene.

\end{document}